\documentclass[prd,aps,nofootinbib,preprint,tightenlines]{revtex4}

\newcommand{\checked}[1]{}

\usepackage{graphicx}
\usepackage{hyperref}

\newcommand{\beq}{\begin{equation}}
\newcommand{\eeq}{\end{equation}}
\newcommand{\bqa}{\begin{eqnarray}}
\newcommand{\eqa}{\end{eqnarray}}

\begin{document}

\title{Parton Energy Loss and Modified Beam Quark Distribution Functions in Drell-Yan Process in $p+A$ Collisions}
\author{Hongxi Xing$^{a,b}$, Yun Guo$^{c,d}$, Enke Wang$^{a}$ and Xin-Nian Wang$^{a,b}$}
\affiliation{ $^a$ Institute of Particle Physics, Central China Normal University, Wuhan 430079, China\\
$^b$Nuclear Science Division, MS 70R0319, Lawrence Berkeley National Laboratory, Berkeley, CA 94720\\
$^c$Department of Physics, Guangxi Normal University, Guilin 541004, China\\
$^d$Department of Physics, Brandon University, Brandon, Manitoba, R7A 6A9 Canada}

\begin{abstract}
Within the framework of generalized collinear factorization in
perturbative QCD (pQCD), we study the effect of initial multiple
parton scattering and induced parton energy loss in Drell-Yan (DY)
process in proton-nucleus collisions. We express the contribution
from multiple parton scattering and induced gluon radiation to the
DY dilepton spectra in terms of nuclear modified effective beam
quark distribution functions. The modification depends on the quark
transport parameter in nuclear medium. This is similar to the
final-state multiple parton scattering in deeply inelastic
scattering (DIS) off large nuclei and leads to the suppression of
the Drell-Yan cross section in $p+A$ relative to $p+p$ collisions.
With the value of quark transport parameter determined from the
nuclear modification of single-inclusive DIS hadron spectra as
measured by the HERMES experiment, we calculate DY spectra in $p+A$
collisions and find the nuclear suppression due to beam parton
energy loss negligible at the Fermilab energy $E_{\rm lab}$=800 GeV
in the kinematic region as covered by the E866 experiment. Most of
the observed nuclear suppression of DY spectra in E866 experiment
can be described well by parton shadowing in target nuclei as given by the EPS08
parameterization. The effect of beam parton energy loss, however,
becomes significant for DY lepton pairs with large beam parton
momentum fraction $x^{\prime}$ or small target parton momentum
fraction $x$. We also predict the DY cross section in $p+A$ collisions 
at lower beam proton energy $E_{\rm lab}$=120 GeV
and show significant suppression due to initial state parton energy loss
at moderately large $x^{\prime}$ where the effect of parton shadowing
is very small.

\end{abstract}
\maketitle

\section{Introduction}
A basic assumption in the collinear factorized parton model in
perturbative QCD (pQCD) is the small intrinsic transverse momentum
of initial partons inside beam hadrons or nuclei relative to both
partons' longitudinal momenta and the large energy-momentum scale
$Q$ involved in hard partonic interactions. To go beyond such a
collinear factorized pQCD model, one has to expand the hard partonic
part of the hard interaction in the intrinsic parton transverse
momentum. In such an expansion, the first term gives rise to the
normal collinear factorized pQCD results, known as the leading twist
contributions which will depend on leading twist
transverse-momentum-integrated parton distribution functions. The
higher order terms in the Taylor expansion can be combined together
with contributions from multiple parton interaction between incoming
or out-going partons and the remanent of target hadrons/nuclei.
These contributions are known as higher twist contributions which
can be expressed as the convolution of hard parts involving multiple
parton scattering and multiple parton correlation functions inside
the target hadrons/nuclei in the framework known as generalized
collinear parton model \cite{Ellis:1982wd,Qiu:1990xxa}. These
higher-twist contributions in general are suppressed by powers of
the momentum scale in the hard processes, $1/Q^{n}$.

In hard processes involving a large nuclear target, higher-twist
contributions often depend on multiple parton correlation functions
inside the target nucleus which are enhanced by the nuclear size
$R_{A}\sim A^{1/3}$ \cite{Luo:1993ui,qgcorrelation,Jorge-Wang}.
These will give rise to a leading and non-trivial nuclear dependance
of the higher-twist contributions to the hard processes beyond the
incoherent superposition of nucleons inside a nucleus.  One example
of such nuclear enhanced higher-twist contributions is the
suppression of leading hadron spectra in deeply inelastic scattering
(DIS) off a nucleus relative to that in DIS off a nucleon and the
suppression is observed to increase with the nuclear size as in the
HERMES experiment \cite{hermes}. Such a suppression has been studied
within the generalized collinear parton model
\cite{Guo:2000nz,Zhang:2003yn,Majumder:2004pt} and was shown to be
caused by multiple parton scattering of the struck quark and induced
gluon radiation inside the target nucleus. The gluon radiation
induced by multiple parton scattering that reduces the energy of the
propagating quark can be expressed as a nuclear modification to the
effective fragmentation function of the struck quark
\cite{Guo:2000nz}, which is determined by the quark-gluon
correlation function inside the nucleus, a higher twist matrix
element of the nucleus not calculable in pQCD. Assuming a large and
loosely bound nucleus, one can approximate these higher twist matrix
elements as a product of initial quark distribution and gluon
density distribution inside the nucleus, the latter is also known as
the quark transport parameter inside the nuclear medium
\cite{qgcorrelation}. The parton transport parameter also describes
the medium broadening of transverse momentum squared
\cite{Majumder:2007hx,Liang:2008vz} per unit distance and is
equivalent to the saturation scale per unit length in a large
nucleus \cite{MV,Kovchegov:1998bi}. One can determine this quark
transport parameter by comparing the leading hadron spectra
suppression in DIS off a nuclear target to experimental data. The
predicted nuclear size and quark energy dependence of the hadron
suppression agree very well with the experiment
\cite{Wang:2002ri,Wang:2009qb}.

One can further extend the study of medium modification of parton
fragmentation functions and parton energy loss or jet quenching due
to multiple parton scattering and induced gluon radiation to the
case of parton propagation inside a hot quark-gluon plasma (QGP) in
heavy-ion collisions. In this case one has to replace the cold
nuclei with a hot thermal medium of quarks and gluons
\cite{Wang:2002ri,Majumder:2007ae,Jorge-Wang}. The parton transport parameter is then
proportional to gluon distribution density inside the hot QGP.
Therefore the study of jet quenching in high-energy heavy-ion
collisions can provide crucial information of space-time profile
of parton density distribution in the early stage of heavy-ion
collisions \cite{GPW}. The observed suppression of single
\cite{ptstar,ptphenix}, dihadron spectra \cite{dihadron} and gamma-hadron
correlation \cite{Adare:2009vd,Abelev:2009gu} in high-energy heavy-ion
collisions at RHIC are all consequences of strong jet quenching and
provided the evidence for the formation of a strongly coupled QGP in
the center of heavy-ion collisions at RHIC \cite{gyumcl,jacwang}.
Recent data from heavy-ion collisions at the LHC \cite{qm11} have
also shown a wide range of phenomena of jet quenching which point to
the formation of a strongly coupled QGP similar to that formed in
heavy-ion collisions at RHIC.

Multiple scattering and induced parton energy loss have been studied
in several theoretical  frameworks \cite{GW,BDMPS, BGZ, GLV}, among
which the higher-twist approach can be applied to jet propagation in
both hot QGP and cold nuclei. In high-energy heavy-ion collisions, energetic partons
which one uses as hard probes have to traverse not only the hot QGP
in the early stage of the dense matter but also hadronic medium
after hadronization of the QGP matter. One therefore has to include the
effect of jet quenching due to multiple parton scattering during the
hadronic phase of the dense matter evolution in high-energy
heavy-ion collisions. Using the information of parton energy loss in
cold nuclei in DIS off large nuclei and extrapolate to hadronic
medium at finite temperature, a recent study \cite{CW} finds
non-negligible contribution from jet quenching in hadronic phase to
the final suppression of large $p_{T}$ hadron spectra in heavy-ion
collisions at RHIC and LHC. Therefore, the study and inclusion of
parton energy loss in cold nuclei and hadronic matter should be an
integral part of phenomenological study of jet quenching in
heavy-ion collisions.

In this paper, we will study the effect of multiple parton
scattering and parton energy loss in the Drell-Yan (DY) process of
dilepton production in $p+A$ collisions. Since dilepton pairs in the
final state do not have strong interaction with the target nucleus,
only initial state interactions are important in DY process.
Therefore, it is an ideal tool to study fast parton propagation and
energy loss in cold nuclear medium
\cite{Gavin:1991qk,ref:E866,Johnson:2001xfa,Arleo:2002ph,Garvey:2002sn,Duan:2008qt,Ivan}.
The higher-twist approach has also been applied to the study of
nuclear effects in the DY process in $p+A$ collisions
\cite{DYHT,Fries}. These studies were restricted to the large
transverse momentum DY pairs, where one can neglect the
interferences associated with induced gluon radiation. In the
calculation of the total DY cross section, one has to integrate over
the transverse momentum of the DY pairs and therefore has to take
into account of the interferences.

Our approach to the nuclear effects in DY pair production in $p+A$ collisions
is very similar to the study of nuclear modification of the effective fragmentation
functions in DIS. Initial multiple parton scattering and induced gluon radiation
before the DY pair production via quark-anti-quark annihilation can lead
to parton energy loss and attenuation of the incoming beam partons.
In the collinear approximation, transverse momentum of the initial
partons are neglected. The transverse momentum of the final DY pairs is
then determined by the transverse momentum of the radiated gluon
either in the vacuum or medium induced gluon bremsstrahlung. In the process involving
multiple parton scattering, there are destructive interferences between the amplitudes
of vacuum gluon bremsstrahlung and gluon radiation induced by multiple scattering.
It is often referred to as the Landau-Pomercanchuk-Migdal (LPM) interference \cite{ref:LPM}.
The LPM interference effect can be characterized by the formation time of the
radiated gluon which is inversely proportional to the transverse momentum squared
$q_{T}^{2}$. For large $q_{T}^{2}$ or short formation time, interference
between vacuum and medium induced gluon bremsstrahlung is negligible. However, for small $q_{T}^{2}$,
the gluon formation time becomes longer than or comparable to the nuclear size and there is
strong destructive interference between vacuum and medium induced gluon bremsstrahlung
that cancels the effect of gluon radiation due to multiple parton scattering.
In our study of DY pair spectra, we will include explicitly the LPM
interference within the framework higher-twist approach.

Similar to the parton energy loss and medium modification of the
effective parton fragmentation functions in DIS, we can express the
nuclear effect in the DY process due to the initial multiple parton
scattering and induced parton energy loss by the beam parton in
terms of modified beam parton distribution functions. The medium
modification can be effectively included in terms of the modified
splitting functions in the QCD evolution equations which depend on
twist-4 quark-gluon correlation functions inside the nucleus.
With the approximation that the quark-gluon parton correlation can
be factorized as a product of the quark and gluon distribution,
the medium modified splitting functions depend only on
the parton transport parameter $\hat{q}$, which is related to the
distribution density of soft gluons inside the nucleus or the
broadening of transverse momentum squared per unit length due to
multiple scattering. This is the only free parameter in the
higher-twist study in both DIS and DY process with a nucleus target.
One therefore can use the information from the phenomenological
study of hadron suppression in DIS experiment \cite{Wang:2009qb} and
calculate the nuclear suppression due to parton energy loss
in the DY process \cite{note}.

Coherence in multiple scattering between beam and target partons
can also lead to the nuclear shadowing of parton distributions
inside the target nucleus. This and other nuclear modifications of
the parton distributions has been measured in nuclear DIS \cite{EMC}
and parametrized \cite{EPS,Hirai:2007sx} from global fitting of experimental data
involving nuclear targets.  These nuclear modifications of parton
distributions can also lead to nuclear effects in DY spectra in
$p+A$ versus that in $p+p$ collisions. We will use the EPS08
parameterization \cite{EPS} of the nuclear parton distribution
functions (nPDF) for the target nucleus in the calculation of DY
spectra in addition to the effect of parton energy loss and medium
modification of the beam parton distributions. We will then compare
the calculated DY spectra to the data from the Fermilab Experiment
866 (E866) \cite{ref:E866}.  We will compare the effect of nuclear
shadowing of target partons and the beam parton energy loss and
investigate in which kinetic region the effect of beam parton energy loss dominate.

The remainder of the paper is organized as the following. In Sec. \ref{gf}, we define
our notations and discuss the leading twist processes and the vacuum
Dokshitzer-Gribov-Lipatov-Altarelli-Parisi (DGLAP) evolution \cite{Dokshitzer:1977sg,Altarelli:1977zs}
of parton distribution functions. In Sec. \ref{tfc}, we calculate the
twist-4 contribution to DY cross section within the framework of generalized factorization.
We then define the effective modified beam quark distribution functions in Sec. \ref{mqdf}
and discuss the energy loss effect on the DY cross section.  We summarize our results in Sec. \ref{sum}.
The complete results of the twist-4 annihilation-like and Compton-like processes are given in
the Appendix \ref{appa} and \ref{appb}, respectively. We also demonstrate that the
twist-4 contributions from double-hard scattering can be understood
as the successive single scatterings in Appendix \ref{appc}.

\section{Leading Twist Contributions}
\label{gf}

In this paper, we focus on the production of DY lepton pairs in unpolarized $h+A$ collisions,
\begin{equation}
h(p')+A(p) \rightarrow \begin{array}{ll}
\gamma^*(q) + X,\\
\;^{|}\!\!\!\rightarrow l^+l^-
\end{array}
\end{equation}
whose cross section can be expressed in terms of the virtual photon $\gamma^{*}(q)$
production cross section,
\begin{eqnarray}
\label{Eq cross section} \frac{d\sigma_{hA \rightarrow
l^+l^-}}{d^4q} =\frac{2\alpha_{em}}{3Q^2} \frac{d\sigma_{hA \rightarrow \gamma^*}}{d^4q}.
\end{eqnarray}
Here $p'$ is the four-momentum of the incoming beam hadron, $p$ is
the momentum per nucleon inside the target nucleus and $q$ is the
four-momentum of the DY lepton pair.
In the center-of-mass frame, we can choose the nucleus and the
beam hadron moving along the ``plus'' and ``minus'' direction, respectively. Using light-cone notations, the
momenta of beam hadron, target nucleus and virtual photon can be expressed as
\begin{eqnarray}
\label{Eq: kinematics}
p'=[0,p'^-,\vec{0}_{\perp}],~~~~~~p=[p^+,0,\vec{0}_{\perp}],~~~~~~q=[(Q^2+q_T^2)/2q^-,q^-,\vec{q}_T],
\end{eqnarray}
where the ``plus'' and ``minus'' components are defined as $p^\pm=(p_0\pm p_z)/\sqrt{2}$.

According to the collinear factorization \cite{factorize}, the Drell-Yan cross section in $h+A$ collisions
can be expressed,
\begin{eqnarray}
\label{Eq cross section}
\frac{d\sigma_{hA \rightarrow \gamma^*}}{d^4q} =\sum\limits_c \int dx^{\prime}
f_{c/h}(x^{\prime}) \frac{d\sigma_{cA \rightarrow \gamma^*}}{d^4q}
\equiv \int dx^{\prime}\frac{d\sigma_{hA \rightarrow \gamma^*}}{d^4qdx^{\prime}},
\end{eqnarray}
in terms of the product of parton$(c)$-nucleus cross section $d\sigma_{cA\rightarrow \gamma^{*}}$
and the normal twist-two beam parton distribution function $f_{c/h}(x^{\prime})$,
where $x'=q^-/p'^-$ is the momentum fraction carried by the annihilated beam quark.

Inside the nuclear target, the beam parton may undergo additional
scattering before the production of the lepton pair. Within the framework of twist
expansion, one can calculate the contribution to the parton-nucleus cross section as a
sum of contributions from single scattering, double scattering and even higher multiple scattering
\begin{eqnarray}
\frac{d\sigma_{cA \rightarrow \gamma^*}}{d^4q}=\frac{d\sigma^S_{cA
\rightarrow \gamma^*}}{d^4q}+\frac{d\sigma^D_{cA \rightarrow
\gamma^*}}{d^4q}+\dots \, .
\end{eqnarray}
We will calculate contributions up to double scatterings in this paper.

\begin{figure}[h]
\begin{center}
\includegraphics[width=5.0cm]{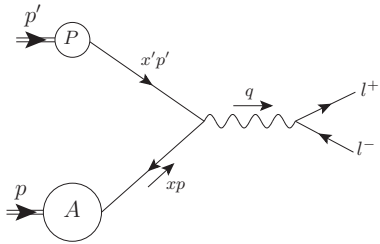}
\centerline{\parbox{10cm}{\caption{\label{Fig-LTLO} \small Lowest
order Drell-Yan process at leading twist.}}}
\end{center}
\end{figure}

\par The leading-twist contribution to DY dilepton pair production at the lowest
order $\mathcal{O}$$(\alpha_s^0)$ comes from the quark anti-quark
annihilation process as illustrated in Fig. \ref{Fig-LTLO}.
The corresponding differential cross section is given by \cite{ref:DY}
\begin{eqnarray}
\label{Eq LO} \frac{d\sigma^{S^{(0)}}_{hA\rightarrow \gamma^{*}}}{dQ^2dx'}= \sum_q \int
dx f_{\bar{q}/A}(x)f_{q/h}(x')H_0(x,p,q)\, ,
\end{eqnarray}
where the summation is over all possible quark and anti-quark and $f_{\bar{q}/A}(x)$ denotes the anti-quark (quark)
distribution function in the nucleus with momentum fraction
 $x$ and the hard part of quark anti-quark annihilation is
\begin{eqnarray}
H_0(x,p,q)=\frac{2\pi\alpha_{em}e_q^2}{3x's}\delta(x-x_B).
\end{eqnarray}
Here $x_B=Q^2/2p\cdot q$ is the Bjorken variable in the DY process, $s=(p+p^{\prime})^2$ is the
center-of-mass energy per $h+N$ collision.

\par
\begin{figure}[h]
\begin{center}
\includegraphics[width=13cm]{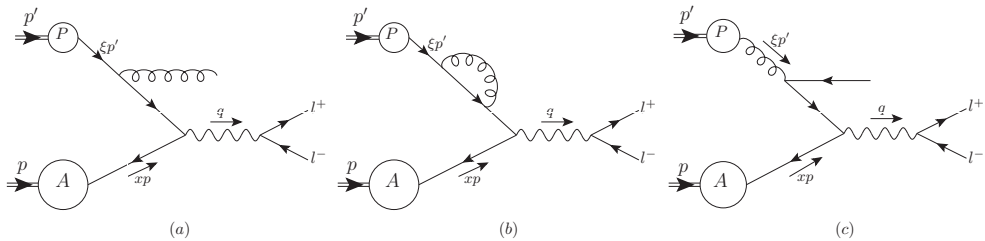}
\centerline{\parbox{15cm}{\caption{\label{Fig-Single}\small
Next-to-leading order Drell-Yan annihilation process (a,b) and
compton scattering (c) at leading twist.}}}
\end{center}
\end{figure}

 To consider the scale dependence of beam quark distribution function, one has to consider
radiative corrections to the annihilation processes and Compton scattering process as illustrated in Fig. \ref{Fig-Single}.
In the annihilation process, the beam quark can radiate a gluon before its annihilation with the anti-quark
from the target as illustrated in Fig. \ref{Fig-Single}(a) and (b).  The annihilating beam quark can also be
generated from a beam gluon splitting as shown in Fig. \ref{Fig-Single}(c) which is actually a Compton
scattering process.

After including the radiative contributions in Fig. \ref{Fig-Single} with the leading logarithmic approximation (for $q_T^2$ up to a
factorization scale $\mu^2$), the renormalized beam quark distribution can be defined as
\begin{eqnarray}
\label{Eq pdf} \nonumber
f_{q/h}(x',\mu^2)&=&f_{q/h}(x')+\frac{\alpha_s}{2\pi}\int_0^{\mu^2}\frac{dq_T^2}{q_T^2}\int_{x'}^{1}
\frac{d\xi}{\xi}\left[f_{q/h}(\xi)\gamma_{q\rightarrow qg}(x'/\xi)\right.\\
&+&\left.f_{g/h}(\xi)\gamma_{g\rightarrow q\bar{q}}(x'/\xi)\right].
\end{eqnarray}
The first and second term inside the square brackets represent contributions from
the annihilation and Compton processes, respectively. The corresponding quark and gluon splitting functions
are given by
\begin{eqnarray}
\label{Eq qq} \gamma_{q\rightarrow
qg}(z)=C_F\left[\frac{1+z^2}{(1-z)_+}+\frac{3}{2}\delta(1-z)\right]\,,
\end{eqnarray}
and
\begin{eqnarray}
\label{Eq gq} \gamma_{g\rightarrow
q\bar{q}}(z)=T_R\left[z^2+(1-z)^2\right]\,,
\end{eqnarray}
where $C_F=4/3$ and $T_R=1/2$ are the color factors.

The renormalized quark distribution function satisfies the vacuum DGLAP
evolution equation,
\begin{eqnarray}
\label{Eq DGLAP} \frac{\partial f_{q/h}(x',\mu^2)}{\partial {\rm
ln}\mu^2}=\frac{\alpha_s}{2\pi}\int_{x'}^{1}
\frac{d\xi}{\xi}\left[f_{q/h}(\xi,\mu^2)\gamma_{q\rightarrow
qg}(x'/\xi)+f_{g/h}(\xi,\mu^2)\gamma_{g\rightarrow
q\bar{q}}(x'/\xi)\right].
\end{eqnarray}
Similarly, one can also obtain the renormalized target quark distribution function and its vacuum DGLAP
evolution equations.

\section{Twist-Four Contribution}
\label{tfc} When a parton is passing through nuclear matter, it
will suffer multiple scattering and parton energy loss. In
principle, contributions from multiple scattering are
power-suppressed for large invariant mass $M=Q$ or
transverse momentum $q_{T}$ of the DY lepton pair. However, if the
involved target partons in multiple scattering come from different
nucleons inside the nucleus, some contributions can be enhanced by the
nuclear size $A^{1/3}$ for each additional scattering. In this
paper, we consider twist-4 contributions which are enhanced by
the nuclear medium size. We will only consider the case of secondary
scattering with another target gluon because the involved
gluon-quark correlation in the nucleus is believed to be much larger
than quark-quark correlation in the double quark rescattering \cite{benweiqq}.

\subsection{Factorization}
\begin{figure}[h]
\includegraphics[width=7.0cm]{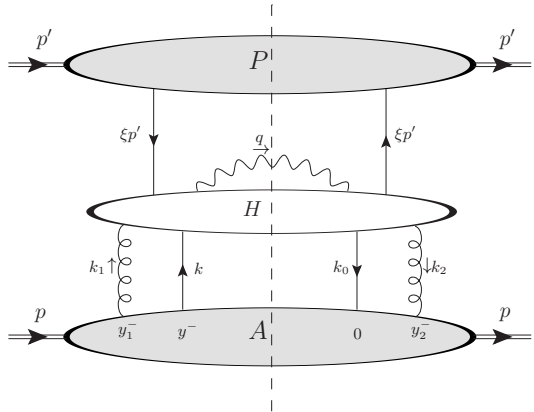}
\centerline{\parbox{15cm}{\caption{\label{Fig-factorization}\small
Next-to-leading order Drell-Yan annihilation process at
twist-four.}}}
\end{figure}

\par In the covariant gauge, the double scattering as shown in Fig. \ref{Fig-factorization} in general
can be cast in the following factorized form:
\begin{eqnarray}
\nonumber
\frac{d\sigma_{qA\rightarrow\gamma^*}^D}{d^4q}&=&\frac{1}{2\xi
s}\int
d^2k_T\int\frac{dy^-}{2\pi}\frac{dy_1^-}{2\pi}\frac{dy_2^-}{2\pi}
\frac{d^2y_T}{(2\pi)^2}e^{-i\vec{k}_T\cdot \vec{y}_T}\overline{H}(y^-,y_1^-,y_2^-,k_T,p,q)\\
&\times&\frac{1}{2}\langle
A|A^+(y_2^-,0_T)\bar{\psi}_q(0)\gamma^+\psi_q(y^-)A^+(y_1^-,y_{T})|A\rangle,
\end{eqnarray}
where $k_0=[(x+x_2)p^+,0,\vec{0}_{\perp}]$,
$k=[xp^+,0,\vec{0}_{\perp}]$, and we keep for now the relative
intrinsic transverse momentum $k_T$ carried by target gluon with
$k_1=[x_1p^+,0,\vec{k}_T]$ and $k_2=[(x_1-x_2)p^+,0,\vec{k}_T]$.
$\overline{H}$ is the Fourier transform of the hard part
$\widetilde{H}$ in momentum space,
\begin{eqnarray}
\nonumber \overline{H}(y^-,y_1^-,y_2^-,k_T,p,q)&=&\int
dxdx_1dx_2e^{ixp^+y^-}e^{ix_1p^+y_1^-}e^{-i(x_1-x_2)p^+y_2^-}\\
&\times&\widetilde{H}(\xi,x,x_1,x_2,k_T,p',p,q).
\end{eqnarray}

As part of the twist expansion,  one makes a Taylor expansion of the partonic
part $\overline{H}$ in the gluon's intrinsic transverse momentum around $k_T=0$ \cite{collinear}:
\begin{eqnarray}
\label{Eq collinear expansion} \nonumber
\overline{H}(y^-,y_1^-,y_2^-,k_T,p,q)&=&\overline{H}(y^-,y_1^-,y_2^-,k_T=0,p,q)\\
&+&\left.\frac{\partial\overline{H}}{\partial{k_T^\alpha}}\right|_{k_T=0}k_T^\alpha
+\frac{1}{2}\left.
\frac{\partial^2\overline{H}}{\partial{k_T^\alpha}\partial{k_T^\beta}}\right|_{k_T=0}k_T^\alpha
k_T^\beta+\dots\,.
\end{eqnarray}
The first term on the right-hand side of the above equation
 can be included in the leading-twist eikonal contribution as the gauge link, which
makes twist-two parton matrix element from the single-scattering gauge invariant.
The second term vanishes after integrating over $k_T$ for
unpolarized initial and final states. The third term will give a
finite contribution to the double-scattering, which, after integrating over $d^2k_T$, becomes
\begin{eqnarray}
\label{Eq twist4 c-section} \nonumber \frac{d\sigma_{qA \rightarrow
\gamma^*}^D}{d^4q}&=&\frac{1}{2\xi s}
\int\frac{dy^-}{2\pi}\frac{dy_1^-}{2\pi}\frac{dy_2^-}{2\pi}
\frac{1}{2}\langle A|F^+_\alpha(y_2^-)\bar{\psi}_q(0)\gamma^+
\psi_q(y^-)F^{+\alpha}(y_1^-)|A\rangle\\
&\times&(-\frac{1}{2}g^{\alpha\beta})\left[\frac{1}{2}\frac{\partial^2}{\partial{k_T^\alpha}
\partial{k_T^\beta}}\overline{H}(y^-,y_1^-,y_2^-,k_T,p,q)\right]_{k_T=0},
\end{eqnarray}
where the factor $k_T^\alpha k_T^\beta$ has been converted
into spatial derivatives in the the field strength tensor after partial integration.

\subsection{Double scattering: Annihilation-like and compton-like processes}
\begin{figure}[h]
\includegraphics[width=15.0cm]{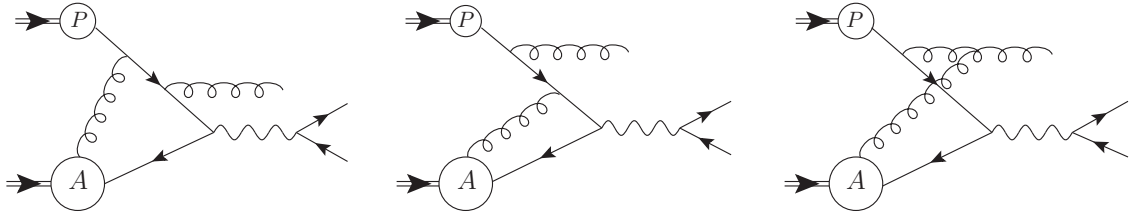}
\centerline{\parbox{15cm}{\caption{\label{Fig-DA}\small
Next-to-leading order Drell-Yan annihilation process at
twist-four.}}}
\end{figure}

There are two distinctive processes at twist-four which we will
consider here. One is the annihilation-like processes as shown in
Fig. \ref{Fig-DA}. Before annihilation into a virtual photon, the
beam quark may scatter with another gluon from the nucleus. Such
multiple parton scattering processes will effectively modify the
distribution of the beam quark. In our calculation, we neglect the
leading order twist-4 processes because our purpose in this paper is
to study the effect of radiative parton energy loss due to the
multiple scattering on the beam quark distribution. Without induced
gluon radiation, the leading order double scattering mainly
contributes to the transverse momentum broadening of the leading
parton \cite{Guo}.
\par
As an illustration of the calculation involved, we consider the contribution from the
first diagram in Fig. \ref{Fig-DA}. Using the Feynman rules,  one can
write down the hard part of this diagram
\begin{eqnarray}
\overline{H}_{A11}=\int
dxdx_1dx_2e^{ixp^+y^-}e^{ix_1p^+y^-_1}e^{-i(x_1-x_2)p^+y^-_2}\frac{1}{2}Tr[\xi\gamma\cdot
p'\hat{H}_{A11}]\Gamma^{(2)},
\end{eqnarray}
where,
\begin{eqnarray}
\nonumber \hat{H}_{A11}&=&\frac{C_F}{2N_c^2}e^2e_q^2g^4
\slash{\hspace{-5pt}p} \frac{\gamma\cdot(\xi p'+k_2)}{(\xi
p'+k_2)^2-i\varepsilon}
\gamma^{\beta}\frac{\gamma\cdot(q-k_0)}{(q-k_0)^2-i\varepsilon}\gamma^{\nu}\frac{1}{2}\gamma\cdot
p\gamma^{\mu}\\
&\times&\frac{\gamma\cdot(q-k)}{(q-k)^2+i\varepsilon}\gamma^{\alpha}
\frac{\gamma\cdot(\xi p'+k_1)}{(\xi
p'+k_1)^2+i\varepsilon}\slash{\hspace{-5pt}p}
(-g_{\mu\nu})\varepsilon_{\alpha\beta}(l_g).
\end{eqnarray}
Throughout this paper, we use the subscript ``Aij'' to denote
different central-cut diagrams at twist-4. We have the following
conventions: ``A'' stands for the annihilation-like
processes while ``i'' and ``j'' correspond to the i-th and j-th
diagrams shown in Fig. \ref{Fig-DA} which are the left and right side of
the cut diagrams, respectively. Similarly, for the Compton-like processes, we use
the subscript``Cij'' to distinguish different central-cut diagrams.

For central-cut
diagrams, the final state phase space is
\begin{eqnarray}
\label{Eq phase space}
\Gamma^{(2)}=\frac{1}{(2\pi)^3}\frac{x_B}{Q^2}\frac{z}{1-z}\delta(x+x_1-x_B-x_t-x_C).
\end{eqnarray}
The momentum fractions are defined as
\begin{eqnarray}
x_t=\frac{x_B}{Q^2}\frac{q_T^2}{1-z},~~~x_C=\frac{x_B}{Q^2}\frac{z(k_T^2-2\vec{q}_{T}\cdot
\vec{k}_T)}{1-z},~~~x_D=\frac{x_B}{Q^2}zk_T^2.
\end{eqnarray}
Performing the contour integration $\int dxdx_1dx_2$, the hard part
becomes
\begin{eqnarray}
\overline{H}_{A11}&=&\alpha_{em}e_q^2\alpha_s^2\frac{C_F}{N_c^2}8(2\pi)^2\frac{1+z^2}{1-z}
\frac{1}{(\vec{q}_T-z\vec{k}_T)^2}\overline{I}_{A11},\\
\nonumber
\overline{I}_{A11}&=&\theta(-y_2^-)\theta(y^--y_1^-)e^{i(x_B+x_t+x_C-x_D)p^+y^-}e^{ix_Dp^+(y_1^--y_2^-)}\\
&\times&\left[1-e^{-i(x_t+x_C-x_D)p^+(y^--y^-_1)}\right]\left[1-e^{-i(x_t+x_C-x_D)p^+y^-_2}\right].
\label{eq:cut1}
\end{eqnarray}
There are four terms in the above result which correspond to the
contributions from four possible physical processes, i.e.,
soft-hard, double-hard and their interferences. This comes out
naturally because of the four possible choices of poles when
performing the contour integrations. The interference are
destructive in small ${q}_T$ region, which cancels contributions
from soft-hard and double-hard processes. In the limit
${q}_T\rightarrow 0$, there is a complete cancelation among all the
processes. This destructive effect is important in our calculation
since we are interested in the transverse momentum integrated
spectra to which the small angle gluon radiation also contributes.

According to the scheme in the twist expansion, the twist-four contributions comes from the
second derivatives of $\overline{H}$ with respect to $k_T$ or the quadratic terms in the Taylor expansion.
In general, one can rewrite the  quadratic term in collinear expansion as
\begin{eqnarray}
\label{Eq expansion} \nonumber
\frac{\partial^2\overline{H}}{\partial k_T^\alpha\partial
k_T^\beta}&\sim& \left[\frac{\partial^2\overline{I}}{\partial
x_{k_T}^2}\frac{\partial x_{k_T}}{\partial k_T^\alpha}\frac{\partial
x_{k_T}}{\partial k_T^\beta}\right]{H}
+\left[\frac{\partial\overline{I}}{\partial
x_{k_T}}\frac{\partial^2 x_{k_T}}{\partial k_T^\alpha\partial
k_T^\beta}\right]{H} + \left[\frac{\partial\overline{I}}{\partial
x_{k_T}}\frac{\partial x_{k_T}}{\partial
k_T^\alpha}\right]\frac{\partial H}{\partial k_T^\beta}\\
&+&\left[\frac{\partial\overline{I}}{\partial x_{k_T}}\frac{\partial
x_{k_T}}{\partial k_T^\beta}\right]\frac{\partial H}{\partial
k_T^\alpha} +\overline{I}\frac{\partial^2 H}{\partial
k_T^\alpha\partial k_T^\beta},
\end{eqnarray}
where $H$ is the truncated hard part of the scattering matrix element when $\bar I$ that
contains all the phase factors is factored out.  We denote $x_{k_T}$ the
$k_T$ dependent fractional momentum, such as $x_C,x_D$ in Eq.~(\ref{eq:cut1}).
As we will show later, the last term of the above equation that contains second derivative
of the truncated hard part $H$ with respect to the intrinsic transverse momentum $k_{T}$
is proportional to a quark-gluon correlation matrix element. Other terms in the
above equations that contain derivatives of the phase factor $I$ will be proportional
to the derivative of the quark-gluon matrix elements with respective to the partons'
momentum fraction. These terms are also proportional to the power of the momentum
fraction $x_{t}\sim q_{T}^{2}/Q^{2}$ and therefore are suppressed relative to the
last term in the above equation for small $q_{T}^{2}/Q^{2}\ll 1$. This feature is
well illustrated by the the second-order derivative of $\overline{H}_{A11}$,
\begin{eqnarray}
\nonumber
\left.-\frac{1}{4}g^{\alpha\beta}\frac{\overline{H}_{A11}}{\partial
k_T^\alpha\partial
k_T^\beta}\right|_{k_T=0}&=&\alpha_{em}e_q^2\alpha_s^2\frac{C_F}{N_c^2}8(2\pi)^2\frac{1+z^2}{1-z}
\frac{1}{q_T^4}\left[z^2\overline{I}_{A11}+z(1-2z)x_t
\frac{\partial\overline{I}_{A11}}{\partial x_C}\right.\\
&+&\left.z(1-z)x_t \frac{\partial\overline{I}_{A11}}{\partial
x_D}+z^2x_t^2 \frac{\partial^2\overline{I}_{A11}}{\partial
x_C^2}\right]_{x_C=0,x_D=0},
\end{eqnarray}
of which the leading contribution in the limit $x_t\sim q_T^2/Q^2\ll 1$ comes
from the first term. However, in the limit $x_{t}p^{+}\ll 1/R_{A}$, LPM interference
becomes important and all the terms are equally important. In this region,
non-perturbative effects also become dominant and the medium modification
cannot be calculable anymore. In our higher-twist approach, we simply assume
the medium modification is still given by the first term in the above equation
that do not involve derivatives of the quark-gluon correlation functions.
Under this assumption, the hard part from the first diagram in Fig. \ref{Fig-DA} is then
\begin{eqnarray}
\nonumber
\left.-\frac{1}{4}g^{\alpha\beta}\frac{\overline{H}_{A11}}{\partial
k_T^\alpha\partial
k_T^\beta}\right|_{k_T=0}&=&\alpha_{em}e_q^2\alpha_s^2\frac{C_F}{N_c^2}8(2\pi)^2\frac{1+z^2}{1-z}
\frac{z^2}{q_T^4}\theta(-y_2^-)\theta(y^--y_1^-)e^{i(x_B+x_t)p^+y^-}\\
&\times&\left[1-e^{-ix_tp^+(y^--y^-_1)}\right]\left[1-e^{-ix_tp^+y^-_2}\right].
\end{eqnarray}

Calculation for the rest of the diagrams and their interferences in Fig.~\ref{Fig-DA} can be similarly carried
out. Summing these contributions together, one can obtain the twist-4 contributions from the annihilation-like
 processes,
\begin{eqnarray}
\label{Eq hard part results1} \nonumber
\left.-\frac{1}{4}g^{\alpha\beta}\frac{\overline{H}_{A}}{\partial
k_T^\alpha\partial
k_T^\beta}\right|_{k_T=0}&=&\alpha_{em}e_q^2\alpha_s^2\frac{1+z^2}{1-z}\frac{8(2\pi)^2}{N_c^2}
\theta(-y_2^-)\theta(y^--y_1^-)e^{i(x_B+x_t)p^+y^-}\\
\nonumber
&\times&\frac{1}{q_T^4}\left[C_Fz^2\left(1-e^{-ix_tp^+(y^--y^-_1)}\right)\left(1-e^{-ix_tp^+y^-_2}\right)\right.\\
\nonumber
&+&C_Fe^{-ix_tp^+(y^--y_1^-)}e^{-ix_tp^+y_2^-}\\
\nonumber
&+&\left(C_F-\frac{C_A}{2}\right)z\left(1-e^{-ix_tp^+(y^--y_1^-)}\right)e^{-ix_tp^+y_2^-}\\
\nonumber
&+&\left(C_F-\frac{C_A}{2}\right)ze^{-ix_tp^+(y^--y_1^-)}\left(1-e^{-ix_tp^+y_2^-}\right)\\
&+&\left.\mathcal{O}(q_T^2/Q^2)\right],
\end{eqnarray}
where the color factor $C_A=3$. The first two terms correspond to
the first two diagrams in Fig. \ref{Fig-DA}, respectively. The
interferences between these two diagrams are given by the third and
fourth term. However, the diagram containing triple-gluon
interaction gives power suppressed contributions $q_{T}^{2}/Q^{2}$
as compared to the other two diagrams and is therefore neglected.
This is quite different from the final state multiple interaction in
the nuclear DIS \cite{Guo:2000nz} where the dominant contribution due to
induced gluon radiation is from the triple-gluon diagram. In
contributions from the first two diagrams there are double-hard,
soft-hard processes for each diagram. Here the second hard interaction
refers to the quark-anti-quark annihilation into DY lepton pair and
the initial interaction with the target gluon can be soft or hard,
depending on the momentum fraction carried by the initial gluon. The
first term has four contributions that include the interferences
between the double-hard and soft-hard processes. For the second
diagram, however, the soft-hard process is prohibited since the
radiated gluon can only
 be induced by the first hard scattering. In this case, only double-hard
process contributes. Notice that one can also obtain the double-hard contributions by
considering two successive parton scatterings in a physical gauge as we demonstrate in
Appendix \ref{appc}. Using the above equation, we can get the leading twist-four
contributions from annihilation-like processes to the DY differential cross section,
\begin{eqnarray}
\label{Eq DAR} \nonumber \frac{d\sigma_{hA \rightarrow
\gamma^*}^{DA(R)}}{dQ^2dx'} &=&\sum_q\int dxH_0(x,p,q)
\int_0^{\mu^2}\frac{dq_T^2}{q_T^4}\frac{\alpha_s}{2\pi}\int_{x'}^1\frac{d\xi}{\xi}f_{q/h}(\xi)
P_{q\rightarrow qg}(x'/\xi)\frac{2\pi\alpha_s}{N_c}\\
&\times&\left[z^2T_{g\bar q}(x,x_t)+T_{g\bar q}^{(1)}(x,x_t)
+\left(1-\frac{C_A}{2C_F}\right)zT_{g\bar q}^{(2)}(x,x_t)\right],
\end{eqnarray}
where the nuclear twist-four matrix elements are defined as
\begin{eqnarray}
\label{Eq matrix element1} \nonumber
T_{g\bar q}(x,x_t)&=&\int\frac{dy^-}{2\pi}dy_1^-dy_2^-e^{i(x+x_t)p^+y^-}
\left(1-e^{-ix_tp^+(y^--y^-_1)}\right)\left(1-e^{-ix_tp^+y^-_2}\right)\\
&\times&\frac{1}{2}\langle
A|F^+_\alpha(y_2^-)\bar{\psi}_q(0)\gamma^+\psi_q(y^-)F^{+\alpha}(y_1^-)|A\rangle
\theta(-y_2^-)\theta(y^--y_1^-),
\end{eqnarray}
\begin{eqnarray}
\label{Eq matrix element2} \nonumber
T_{g\bar q}^{(1)}(x,x_t)&=&\int\frac{dy^-}{2\pi}dy_1^-dy_2^-e^{ixp^+y^-}e^{ix_tp^+(y_1^--y_2^-)}\\
&\times&\frac{1}{2}\langle
A|F^+_\alpha(y_2^-)\bar{\psi}_q(0)\gamma^+\psi_q(y^-)F^{+\alpha}(y_1^-)|A\rangle
\theta(-y_2^-)\theta(y^--y_1^-),
\end{eqnarray}
\begin{eqnarray}
\label{Eq matrix element3} \nonumber
T_{g\bar q}^{(2)}(x,x_t)&=&\int\frac{dy^-}{2\pi}dy_1^-dy_2^-e^{i(x+x_t)p^+y^-}
\left[e^{-ix_tp^+y_2^-}-2e^{-ix_tp^+(y^--y_1^-+y_2^-)}+e^{-ix_tp^+(y^--y_1^-)}\right]\\
&\times&\frac{1}{2}\langle
A|F^+_\alpha(y_2^-)\bar{\psi}_q(0)\gamma^+\psi_q(y^-)F^{+\alpha}(y_1^-)|A\rangle
\theta(-y_2^-)\theta(y^--y_1^-).
\end{eqnarray}
Here $T_{g\bar q}$ and $T_{g\bar q}^{(1)}$ are related to the first
and second diagram in Fig. \ref{Fig-DA}, respectively, and $T_{g\bar
q}^{(2)}$ arises from the interference between them. Note that the
interference between the soft-hard and double hard processes depend
on the nuclear size and the momentum fraction $x_t$ from either the
target quark or gluon that is needed for the induced gluon
radiation. The interference can be neglected when
$x_{t}p^{+}R_{A}\gg 1$ for  large values of  $q_T$ or short
formation length for the gluon radiation relative to the nuclear
size.  In the small $q_{T}$ region or large formation time, however,
the interference effect becomes important due to what is referred to
as LPM effect \cite{ref:LPM}. The destructive interference is
dictated by the gluon formation time, $\tau_f\equiv 1/x_tp^+$,
which, in the collinear limit, could become much larger than the
relative distance between the two scattering centers. Therefore the
LPM effect  will effectively cut off the transverse momentum of the
virtual photon at $q_T^2\sim Q^2/R_A$ \cite{Guo:2000nz} in the
leading calculation of the DY lepton pair production. This will lead
to an anomalous nuclear dependence of the DY cross section in $h+A$
collisions.

\par We also need to consider the interferences between
single and triple scattering. Our calculations
show that all the contributions from these processes are
power suppressed because the leading term, i.e., the last term in Eq.
(\ref{Eq expansion}), vanishes. However, when combining all the
contributions from both the double scattering and single-triple
interferences, the first term in the collinear expansion  in Eq.
(\ref{Eq collinear expansion}) gives the
eikonal contributions to the single scattering. Such contributions
do not correspond to any physical scattering since
they can be gauged away, but ensures the gauge invariance of the leading
twist results. We list the full results of the single-triple interferences in the appendix.

Similarly to the leading twist case, one also needs to consider the virtual contribution
which corresponds to the quark self-energy correction. The results of virtual
correction can be obtained by the requirement of unitarity.
\begin{eqnarray}
\nonumber \frac{d\sigma_{hA \rightarrow \gamma^{*}}^{DA(V)}}{dQ^2dx'}
&=&-\sum_q\int dxH_0(x,p,q)
\int_0^{\mu^2}\frac{dq_T^2}{q_T^4}\frac{\alpha_s}{2\pi}\int_{0}^1dzf_{q/h}(x')
P_{q\rightarrow qg}(z)\frac{2\pi\alpha_s}{N_c}\\
&\times&\left[z^2T_{g\bar q}(x,x_t)+T_{g\bar q}^{(1)}(x,x_t)
+\left(1-\frac{C_A}{2C_F}\right)zT_{g\bar q}^{(2)}(x,x_t)\right].
\end{eqnarray}
Including both radiative and virtual contributions, the final result for the
annihilation-like process is given by
\begin{eqnarray}
\label{Eq DA} \frac{d\sigma_{hA \rightarrow \gamma^{*}}^{DA}}{dQ^2dx'}
&=&\sum_q\int dxf_{\bar{q}/A}(x)
\int_0^{\mu^2}\frac{dq_T^2}{q_T^2}\frac{\alpha_s}{2\pi}\int_{x'}^1\frac{d\xi}{\xi}f_{q/h}(\xi)
\Delta\gamma_{q\rightarrow qg}(x'/\xi,q_T^2)H_0(x,p,q)\, .
\end{eqnarray}
Here, the modified quark splitting function is
\begin{eqnarray}
\label{Eq msplitting qq} \Delta\gamma_{q\rightarrow
qg}(z,q_T^2)=\left[\frac{1+z^2}{(1-z)_+}T_{g\bar
q}^{A}+\delta(1-z)\Delta T_{g\bar
q}^{A}\right]\frac{C_F2\pi\alpha_s}{q_T^2N_cf_{\bar{q}/A}(x)}
\end{eqnarray}
with
\begin{eqnarray}
T_{g\bar q}^{A}=z^2T_{g\bar q}(x,x_t)+T_{g\bar q}^{(1)}(x,x_t)
+\left(1-\frac{C_A}{2C_F}\right)zT_{g\bar q}^{(2)}(x,x_t)
\end{eqnarray}
and
\begin{eqnarray}
\Delta T_{g\bar q}^{A}\equiv\int_0^1
dz\frac{1}{1-z}\left[2T_{g\bar q}^{A}|_{z=1} -(1+z^2)T_{g\bar q}^{A}\right].
\end{eqnarray}
Notice that the infrared divergence is canceled between the real and
virtual corrections in the same way as in the single scattering.

\begin{figure}[h]
\includegraphics[width=15.0cm]{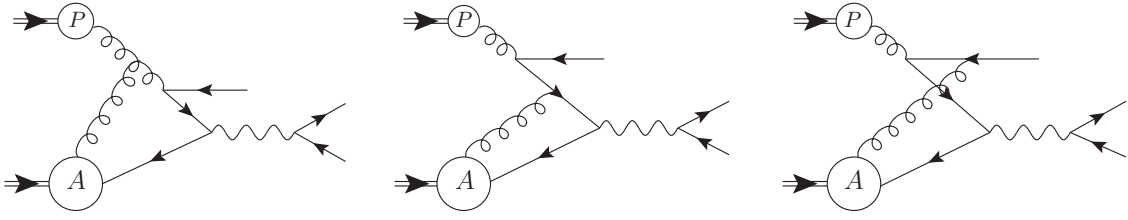}
\centerline{\parbox{15cm}{\caption{\label{Fig-DC}\small
Next-to-leading order Drell-Yan compton process at twist-four.}}}
\end{figure}

\par The other twist-four processes that contribute to the modified beam quark distribution function
are the Compton-like processes as shown in Fig. \ref{Fig-DC}. The full results of these
 processes can be found in the appendix. Similarly to the calculation of annihilation-like
processes, one can get the leading twist-4 contributions,
\begin{eqnarray}
\label{Eq HDC} \nonumber
\left.-\frac{1}{4}g^{\alpha\beta}\frac{\overline{H}_{C}}{\partial
k_T^\alpha\partial
k_T^\beta}\right|_{k_T=0}&=&\alpha_{em}e_q^2\alpha_s^2(2z^2-2z+1)\frac{8(2\pi)^2}{N_c(N_c^2-1)}
e^{i(x_B+x_t)p^+y^-}\theta(-y_2^-)\theta(y^--y_1^-)\\
\nonumber
&\times&\frac{1}{q_T^4}\left[C_Az^2\left(1-e^{-ix_tp^+(y^--y^-_1)}\right)\left(1-e^{-ix_tp^+y^-_2}\right)\right.\\
\nonumber
&+&C_Fe^{-ix_tp^+(y^--y_1^-)}e^{-ix_tp^+y_2^-}+\frac{C_A}{2}z\left(1-e^{-ix_tp^+(y^--y_1^-)}\right)e^{-ix_tp^+y_2^-}\\
&+&\left.\frac{C_A}{2}ze^{-ix_tp^+(y^--y_1^-)}\left(1-e^{-ix_tp^+y_2^-}\right)+\mathcal{O}(q_T^2/Q^2)\right].
\end{eqnarray}
which are analogous to the results from the annihilation-like processes.
The first two terms correspond to the first two diagrams in Fig. \ref{Fig-DC}, respectively.
The third and fourth terms are the interference contributions between the two diagrams.
The contribution from the last diagram in Fig. \ref{Fig-DC} is neglected as its contribution
is power suppressed. Again, one can obtain the double-hard contributions by considering
two successive single scatterings as shown in Appendix \ref{appc}. Using the above results,
for  Compton-like process, Eq. (\ref{Eq twist4 c-section}) finally reads

\begin{eqnarray}
\label{Eq DC} \frac{d\sigma_{hA \rightarrow l^+l^-}^{DC}}{dQ^2dx'}
&=&\sum_q\int dxf_{\bar{q}/A}(x)
\int_0^{\mu^2}\frac{dq_T^2}{q_T^2}\frac{\alpha_s}{2\pi
}\int_{x'}^1\frac{d\xi}{\xi}f_{g/h}(\xi) \Delta\gamma_{g\rightarrow
q\bar{q}}(x'/\xi,q_T^2)H_0(x,p,q),
\end{eqnarray}
where the modified gluon splitting function is defined as
\begin{eqnarray}
\label{Eq msplitting gq} \Delta\gamma_{g\rightarrow
q\bar{q}}(z,q_T^2)=\left[(1-z)^2+z^2\right]T_{g\bar
q}^{C}\frac{2\pi\alpha_sC_A}{q_T^2(N_c^2-1) f_{\bar{q}/A}(x)},
\end{eqnarray}
with the twist-four matrix element
\begin{eqnarray}
T_{g\bar q}^{C}=z^2T_{g\bar q}(x,x_t)+\frac{C_F}{C_A}T_{g\bar
q}^{(1)}(x,x_t) +\frac{1}{2}zT_{g\bar q}^{(2)}(x,x_t).
\end{eqnarray}
Diagrams in Fig.
\ref{Fig-DC} contributing to each term in the above equation are similar
to the annihilation-like processes. Here, the
twist-four matrix elements are the same as in the annihilation-like
processes as defined in Eq. (\ref{Eq matrix element1})-(\ref{Eq
matrix element3}). In Compton-like
processes, there is no virtual correction at $\mathcal{O}(\alpha_s)$,
and the gluon splitting function is infrared finite.

\section{Modified quark distribution function}
\label{mqdf}

Summing up all the leading logarithmic contributions
from single [Eq. (\ref{Eq pdf})] and double scattering [Eqs.
(\ref{Eq DA})  and (\ref{Eq DC})], we can define the medium-modified
effective beam quark distribution function as
\begin{eqnarray}
\label{Eq mpdf} \nonumber
\tilde{f}_{q/h}(x',\mu^2,A)&=&f_{q/h}(x',\mu^2)+\frac{\alpha_s}{2\pi}\int_0^{\mu^2}\frac{dq_T^2}{q_T^2}
\int_{x'}^{1}\frac{d\xi}{\xi}\left[f_{q/h}(\xi)\Delta\gamma_{q\rightarrow qg}(x'/\xi,q_T^2)\right.\\
&+&\left.f_{g/h}(\xi)\Delta\gamma_{g\rightarrow
q\bar{q}}(x'/\xi,q_T^2)\right]\, ,
\end{eqnarray}
where $f_{q/h}(x',\mu^2)$ is the  renormalized twist-two beam quark
distribution given in Eq.~(\ref{Eq pdf}). The $A$-dependence of the medium-modified beam
quark distribution function is implicit through the medium-modified splitting functions in the
second term on the right-hand side. Assuming that multiple gluon
bremsstrahlung induced by additional scattering in the nuclear medium can be resummed in the same way as in the vacuum, one
can obtain the medium-modified DGLAP evolution equation for the beam quark
distribution function
\begin{eqnarray}
\label{mdglap} \frac{\partial\tilde{f}_{q/h}(x',\mu^2,A)}{\partial
ln\mu^2}=\frac{\alpha_s}{2\pi}\int_{x'}^{1}
\frac{d\xi}{\xi}\left[\tilde{f}_{q/h}(\xi,\mu^2)\tilde{\gamma}_{q\rightarrow
qg}(x'/\xi,\mu^2)+\tilde{f}_{g/h}(\xi,\mu^2)\tilde{\gamma}_{g\rightarrow
q\bar{q}}(x'/\xi,\mu^2)\right],
\end{eqnarray}
where the modified splitting functions,
\begin{eqnarray}
\label{final split}
\tilde{\gamma}_{a\rightarrow bc}(z,\mu^2)=\gamma_{a\rightarrow
bc}(z)+\Delta\gamma_{a\rightarrow bc}(z,\mu^2),
\end{eqnarray}
are the sum of contributions from both the vacuum and the nuclear medium-induced
bremsstrahlung. Notice that Eq. (\ref{mdglap}) is formally the same as the DGLAP
evolution equation in vacuum, Eq. (\ref{Eq DGLAP}), but its
splitting functions have extra terms $\Delta\gamma$ which are caused by
medium-induced gluon radiation. Such medium-induced terms will
soften the beam quark distribution function as a consequences of
radiative energy loss by the leading beam parton.

\par
\begin{figure}[h]
\begin{center}
\includegraphics[width=7.0cm]{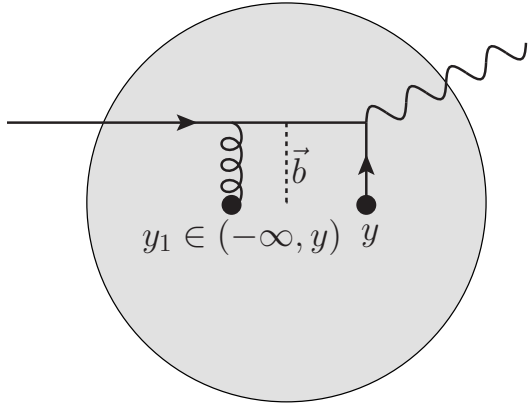}
\centerline{\parbox{10cm}{\caption{\label{Fig-path}\small
Illustration of PA scattering.}}}
\end{center}
\end{figure}

 As we mentioned before, the medium-modified splitting functions
$\Delta\gamma$ depend on the twist-four matrix elements
$T_{g\bar{q}}^{A}$. In order to get a numerical estimate of the
effect of initial parton energy loss, we need to know
$T_{g\bar{q}}^{A}$ which is not calculable. Neglecting correlation
between the quark and gluon distribution, one can assume the
factorization of the twist-four matrix elements \cite{Wang:2009qb}
in terms of quark and gluon distributions \cite{qgcorrelation} from
two independent nucleons inside the nucleus, the later in turn can
be related to quark transport parameter $\hat{q}$ in nuclear medium.
Under these assumptions, the twist-four matrix elements  can be
expressed as
\begin{eqnarray}
\label{tmatrix} T_{g\bar q}(x,x_t)&=&2
\pi\int_{-\infty}^{\infty}dy^- \int_0^{\infty} d^2 {b}
\int_{-\infty}^{y^-}dy_1^-\rho_A(y_1^-,\vec{b})\rho_A(y^-,\vec{b}) \nonumber \\
&\times&{\rm sin}^2(x_tp^+y_1^-/2)
f_{\bar{q}/N}^{A}(x,\vec{b})\left[x
f_{g/N}^{A}(x,\vec{b})|_{x\approx 0}+x_t
f_{g/N}^{A}(x_t,\vec{b})\right],
\end{eqnarray}
where, $f_{a/N}^{A}(x,\vec{b})$ is the parton distribution function
per nucleon inside the nucleus. As illustrated in
Fig.~\ref{Fig-path}, $\vec{b}$ is the impact parameter of $h+A$
collisions, $y_{1}^{-}$ is the longitudinal position of the target
nucleon where the soft gluon comes from and $y^{-}$ is the
longitudinal position of the target nucleon where the final
quark-antiquark annihilation into DY lepton pair takes place. One
has to integrate over the impact parameter $\vec{b}$. The nucleon
density distribution $\rho_A(y^-,\vec{b})$ is normalized as
\begin{eqnarray}
A=\int_{-\infty}^{\infty}dy^- \int_0^{\infty} d^2
{b}\,\rho_A(y^-,\vec{b}) . \label{eq:norm}
\end{eqnarray}

So far we have focused on the nuclear modification of beam quark
distribution function in DY processes in $h+A$ collisions due to
multiple parton scattering and medium-induced parton energy loss.
Multiple parton interaction and their coherence can also lead to
nuclear modification of the target parton distributions. Gluon
fusion in the evolution of the target parton distribution function
can also lead to parton saturation \cite{QM}. Some of these effects
are higher-twist as we have discussed so far and some are leading
twist which are determined by the wave function of the nucleus and
initial parton distribution functions. In this paper, we will adopt
a phenomenological approach for the nuclear modification of the
target parton distributions and use the current parameterization of
the nuclear parton distribution functions (nPDF) from global fit to
the existing experimental data. In this case, the effective target
parton distribution function per nucleon at impact parameter
$\vec{b}$ is defined as
\begin{equation}
f_{a/N}^{A}(x,\vec{b}) \equiv \frac{1}{A} f_{a/A}(x,\vec{b}) \equiv
R_{A}^{a}(x,\vec{b}) f_{a/N}(x),
\end{equation}
where $f_{a/N}(x)$ is the nucleon parton distribution function in
vacuum and $R_{A}^{a}(x,\vec{b})$ is the nuclear modification factor
for parton distribution functions inside a nucleus at impact
parameter $\vec{b}$. The impact parameter averaged parton
distribution function in a nucleus is defined as
\begin{equation}
f_{a/A}(x)\equiv f_{a/N}(x) \int d^{2}{b} t_A(\vec{b})
R_{A}^{a}(x,\vec{b}),
\end{equation}
where $t_A(\vec{b})=\int dy^{-}\rho_A(y^-,\vec{b})$ is the nuclear
thickness function. Note that the nuclear modification factor
$R_{A}^{a}(x,\vec{b})$ should contain the information about the
isospin of the nucleus.

We further define the generalized jet transport parameter as
\begin{eqnarray}
\hat{q}(x_t,y^-,\vec{b})&\equiv &\frac{4\pi^2\alpha_s
C_R}{N_c^2-1}\rho_A(y^-,\vec{b})\frac{1}{2} \left[x
f_{g/N}^{A}(x,\vec{b})|_{x\approx
0}+x_t f_{g/N}^{A}(x_t,\vec{b})\right] \nonumber \\
&\approx & \hat q_{0}
\frac{\rho_A(y^-,\vec{b})}{\rho_A(0,\vec{0}_{\perp})},
\end{eqnarray}
assuming the limit $x_t \ll 1$, where $\hat q_{0}$ is the quark transport parameter in the center of the
nucleus. With the above definition of quark transport parameter and nuclear quark distribution
function, the twist-four matrix elements can be expressed as,
\begin{eqnarray}
T_{g\bar q}(x,x_t)&\approx&\frac{6\hat{q}_0 f_{\bar
q/N}(x)}{\pi\alpha_s \rho_A(0,\vec{0}_{\perp})}\int
d^2{b}\int_{-\infty}^{\infty}dy^-\int_{-\infty}^{y^-}
dy_1^- \rho_A(y_1^-,\vec{b})\rho_A(y^-,\vec{b}) \nonumber \\
&\times& {\rm sin}^2(x_tp^+y_1^-/2) R_{A}^{\bar q}(x,\vec{b}).
\end{eqnarray}

In our numerical evaluation, we will
use the Woods-Saxon distribution for the nucleon distribution
\begin{eqnarray}
\rho_A(y,\vec{b})=\frac{\rho_0}{1+\exp(\frac{r-R_A}{a})},
\end{eqnarray}
where $r=\sqrt{{b}^2+y^2}$, $R_A\approx 1.12 A^{1/3}$ fm is the
radius of the nucleus, $a \approx 0.537 $ the width of the nuclear
skin and the constant $\rho_0$ is
 fixed by normalization in Eq.~(\ref{eq:norm}).

\par Similarly, the other two matrix elements in Eqs. (\ref{Eq matrix
element2}) and (\ref{Eq matrix element3}) can be approximated as
\begin{eqnarray}
T_{g\bar q}^{(1)}(x_B,x_t)&\approx &\frac{3\hat{q}_0 f_{\bar
q/N}(x)}{2\pi\alpha_s \rho_A(0,\vec{0}_{\perp})}\int
d^2{b}\int_{-\infty}^{\infty}dy^-\int_{-\infty}^{y^-} dy_1^-
\rho_A(y_1^-,\vec{b})\rho_A(y^-,\vec{b}) R_{A}^{\bar q}(x, \vec{b}),\\
T_{g\bar q}^{(2)}(x_B,x_t)&\approx &\frac{-6\hat{q}_0 f_{\bar
q/N}(x)}{\pi\alpha_s \rho_A(0,\vec{0}_{\perp})}\int
d^2{b}\int_{-\infty}^{\infty}dy^-\int_{-\infty}^{y^-} dy_1^-
\rho_A(y_1^-,\vec{b})\rho_A(y^-,\vec{b}) \nonumber \\
&\times& {\rm sin}^2(x_tp^+y_1^-/2) R_{A}^{\bar q}(x, \vec{b}) .
\end{eqnarray}

Because of the LPM interference effect, the formation time for the
induced parton radiation cannot be larger than the nuclear size.
This will restrict the transverse momentum $q_T$ to a limited region
$q_T^2\sim Q^2/A^{1/3}$. Below this limit, the contributions will be
suppressed because of the destructive LPM interference. Such LPM
effect in contributions that contain  $T_{g\bar q}$ and $T_{g\bar
q}^{(2)}$ give rise to an addition factor of $A^{1/3}$ after
integration  over the transverse momentum. This nuclear enhancement
leads to the quadratic nuclear size dependence of the parton energy
loss. Relative to these two contributions and in the large nucleus
limit, we can neglect contributions proportional to $T_{g\bar
q}^{(1)}$ which does  not contain LPM effect \cite{Zhang:2003yn}.

\begin{figure}[h]
\begin{center}
\includegraphics[width=15.0cm]{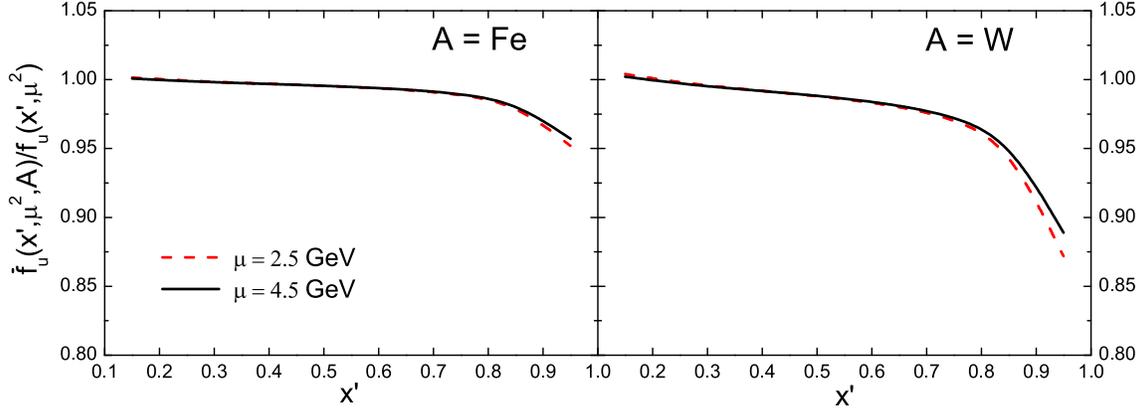}
\centerline{\parbox{15cm}{\caption{\label{Fig-quarkdist}\small
Nuclear modification of the effective beam u-quark distribution
functions $\tilde{f}_u(x',\mu^2,A)/f_u(x',\mu^2)$ versus $x'$ in DY
processes in $p+A$ collisions at $E_{lab}=800$ GeV and scale
$\mu=4.5$ (solid) and $2.5$ GeV (dashed). The quark transport
parameter is set at $\hat{q}_0=0.024~\rm
GeV^2/fm$\cite{Wang:2009qb}.}}}
\end{center}
\end{figure}

Shown in Fig.~\ref{Fig-quarkdist} are the effective nuclear
modification factors of the beam u-quark distributions in $p+W$ and
$p+Fe$ DY processes at $E_{\rm lab}=800$ GeV. In this numerical
calculation and in the rest of this paper, we will only consider the
effective nuclear modified beam parton distributions due to double
scattering as given in Eq.~(\ref{Eq mpdf}). The normal twist-2 quark
distributions in nucleon are given by the CTEQ6L \cite{ref:CTEQ}
parameterization, the nuclear modification of parton distribution
functions inside a nuclear target will be given by the EPS08
\cite{EPS} parameterization without impact-parameter dependence. As
shown in the figure, the nuclear modification of the beam quark
distribution function is negligible in small $x'$ region. The
modification becomes significant in the large $x'$ region for small
factorization scale $\mu$ and large target nuclei. We also observe
some small scale dependence of the nuclear modification.

\par Experimentally, the Drell-Yan differential cross section can be measured
for different targets in hadron-nucleus collisions. The cross section ratios are
analyzed for the study of nuclear effects,
\begin{eqnarray}
\label{ratio}
\frac{B \sigma^A}{A \sigma^B}=\frac{B d \sigma_{pA\rightarrow
l^+l^-}/dQ^2dx'}{ A d\sigma_{pB\rightarrow l^+l^-}/dQ^2dx'}.
\end{eqnarray}
In general, one would like to use the experimental data in $p+p$ collisions as the
reference in the denominator to study the nuclear effects in $p+A$ collisions with a large
nucleus target.  One, however, often uses $p+B$ with a light nucleus target $B$
as an approximate of the $p+p$ reference data (modulo isospin corrections)
as in the Fermilab E866 \cite{ref:E866} experiment.  Under this approximation,
the measured nuclear modification can be compared to
\begin{eqnarray}
\frac{B \sigma^A}{A \sigma^B}=\frac{\sum_q \int dx
f_{\bar{q}/A}(x,\mu^2)\tilde{f}_{q/p}(x',\mu^2,A)H_0(x,p,q)}{A \sum_q
\int dx f_{\bar{q}/N}(x,\mu^2) f_{q/p}(x',\mu^2)H_0(x,p,q)},
\end{eqnarray}
where $\tilde{f}_{q/p}(x',\mu^2,A)$ is the modified beam quark
distribution function defined in Eq. (\ref{Eq mpdf}), and
$f_{\bar{q}/N}(x,\mu^2)$ is the normal twist-2 anti-quark
distribution in nucleon (with the isospin of the nucleus $B$) which
will be given by the CTEQ6L \cite{ref:CTEQ} parameterization in our
study here. As for the nuclear modification of parton distribution
functions inside a nuclear target, we will use the EPS08 \cite{EPS}
parameterization without impact-parameter dependence.

\begin{figure}[h]
\begin{center}
\includegraphics[width=15.0cm]{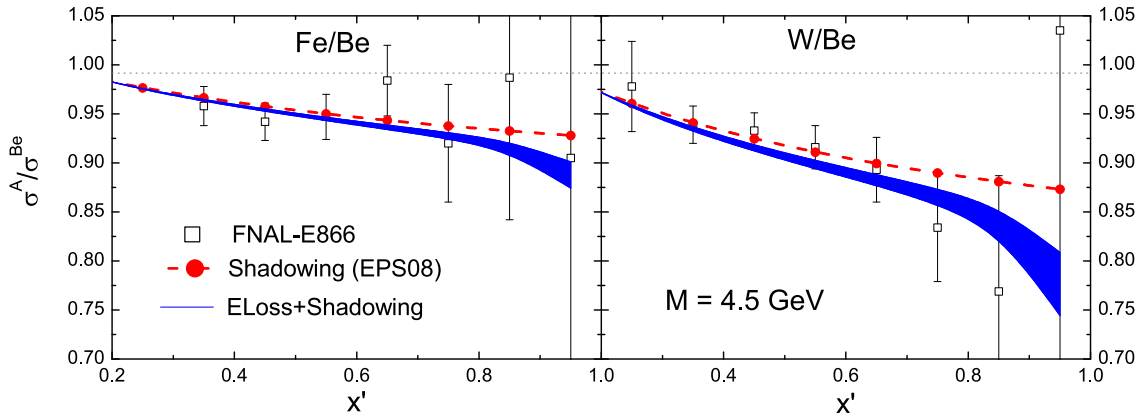}
\centerline{\parbox{15cm}{\caption{\label{Fig-ratio}\small Ratios of
the DY cross section per nucleon in $p+A$ collisions versus $x'$ for
Fe/Be and W/Be at $E_{lab}=800$ GeV and invariant dilepton mass
$M=4.5$ GeV. The curves are theoretical results with (solid) and
without (dashed) initial beam parton energy loss. The shaded bands
correspond to $\hat{q}_0=0.024\pm0.008~\rm GeV^2/fm$
\cite{Wang:2009qb}. The experimental data are from the Fermilab
experiment E866 \cite{ref:E866}. }}}
\end{center}
\end{figure}

Shown in Fig.~\ref{Fig-ratio} are the calculated DY cross section
ratios as a function of $x'$ for different nuclei ($Be$, $Fe$ and $W$)
with (solid) and without (dashed) the effect of initial beam parton
energy loss via multiple interaction, as compared to the Fermilab E866
experimental data \cite{ref:E866}. In
our numerical evaluation of the effect of initial beam parton energy
loss,  the quark transport parameter $\hat{q}_0$ at the center of a large
nucleus is the only free parameter and is fixed to
$\hat{q}_0=0.024\pm0.008\, {\rm GeV^2/fm}$
from the phenomenological study of parton energy loss and
modified fragmentation function in nuclear DIS by the HERMES
experiment \cite{Wang:2009qb}. As shown in the comparison between our calculation
and the experimental data in the kinematic region of the E866
experiment, the dominant nuclear modification of the DY cross
section is from nuclear shadowing of parton distribution functions
inside large nuclei as given by the EPS08 \cite{EPS}
parameterization. For fixed invariant mass $M$, the fractional
momentum $x=M^{2}/x^{\prime}s$ carried by the target partons becomes
smaller for large beam quark fractional momentum $x^{\prime}$,
therefore strong nuclear shadowing of the quark distribution inside
the target nucleus. The effect of medium-modified beam quark
distribution caused by beam quark energy loss leads to further
suppression of the DY cross section for large nuclei. However, with
the quark transport parameter predetermined from the nuclear DIS
experiment \cite{Wang:2009qb}, the suppression due to initial beam
quark energy loss is quantitatively small. This is consistent
with other estimates of parton energy loss in DY process
\cite{Johnson:2001xfa,Arleo:2002ph,Garvey:2002sn}. The additional suppression only becomes
considerable in large $x'$ region in a large nucleus where one also
see large nuclear modification of the effective beam quark
distribution function as shown in Fig. \ref{Fig-quarkdist}.  Since
the parameterization of nPDF \cite{EPS} from global fitting included
DY data, one should include the effect of beam parton energy loss
in large $x'$ or small $x$ region.

\begin{figure}[h]
\begin{center}
\includegraphics[width=15.0cm]{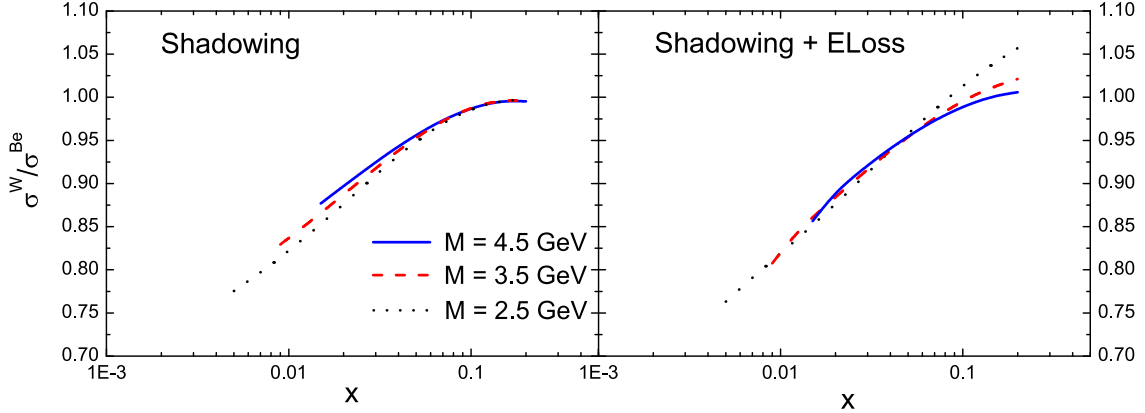}
\centerline{\parbox{15cm}{\caption{\label{Fig-scaling}\small Ratios
of the DY cross section per nucleon in $p+W$ and $p+Be$ collisions
versus $x$ at $E_{lab}=800$ GeV for invariant dilepton mass $M=4.5$
(solid), 3.5 (dashed) and 2.5 (doted) GeV with only parton shadowing
(left panel) and shadowing and initial beam parton energy loss
(right panel). The quark transport parameter is set at
$\hat{q}_0=0.024 ~{\rm GeV}^2/fm$ \cite{Wang:2009qb}.}}}
\end{center}
\end{figure}

Without initial beam parton energy loss, the nuclear modification of
the DY cross section as a function of the target parton momentum
fraction $x$ should more or less reflect that of the parton
shadowing, which is shown in the left panel of
Fig.~\ref{Fig-scaling}. The slight $x$-scaling violation for
different invariant DY dilepton mass $M$ is caused mainly by the
scale violation of the parton shadowing effect as given by the EPS08
\cite{EPS} parameterization. Parton shadowing becomes smaller for
larger invariant mass due to the QCD evolution of the nuclear parton
distributions which leads to less suppression of the DY cross
section. When one takes into account of the beam parton energy loss,
there is increased suppression of DY cross section at large
$x^{\prime}$ or small $x$ for fixed invariant mass $M$ as shown in
the right panel of Fig.~\ref{Fig-scaling}. For fixed values of $x$,
an increase of invariant mass $M$ leads to an increase in
$x^{\prime}$. This should lead in turn an increased suppression due
to beam parton energy loss according to Figs.~\ref{Fig-ratio}. Such
increased suppression due to larger values of $x^{\prime}$ will
counter the effect of scale violation of the shadowing (see the left
panel). The final outcome is an approximate $x$-scaling of the
suppression factor in small $x$ region for different values of the
DY dilepton mass.

\begin{figure}[h]
\begin{center}
\includegraphics[width=15.0cm]{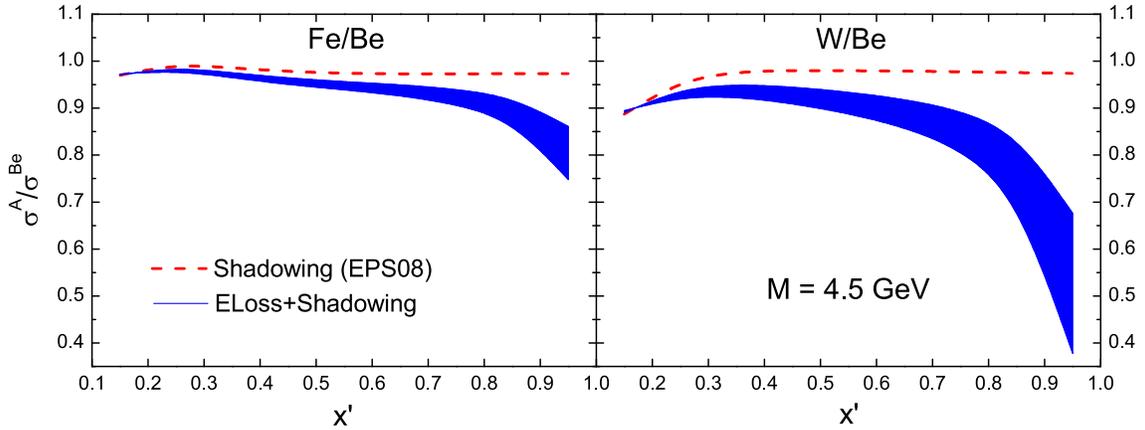}
\centerline{\parbox{15cm}{\caption{\label{Fig-e906}\small
Predictions for the DY cross section ratios per nucleon in $p+A$
collisions versus $x'$ for Fe/Be and W/Be at
($E_{lab}=120$ GeV) with invariant dilepton mass $M=4.5$ GeV. The
curves are theoretical results with (solid) and without (dashed)
initial beam parton energy loss. The shaded bands correspond to
$\hat{q}_0=0.024\pm0.008~\rm GeV^2/fm$ \cite{Wang:2009qb}.}}}
\end{center}
\end{figure}

Nuclear shadowing of parton distributions in the target nucleus is significant only at small
momentum fraction $x$. For fixed invariant mass of DY lepton pairs and moderately large beam
parton momentum fraction $x^{\prime}$ or large target parton momentum fraction $x$, the effect of shadowing
should be small.  On the other hand the fractional parton energy loss will become larger for smaller
beam parton energy $x^{\prime}E_{lab}$ \cite{Wang:2009qb}. Therefore, at fixed DY dilepton mass $M$ and 
moderately large beam parton momentum fraction $x^{\prime}$, the effect of the parton energy loss on
DY cross section in $p+A$ collisions will become more dominant for lower beam energy $E_{lab}$.
This will enable one to  disentangle the effect of initial-state energy loss
effect from the nuclear shadowing. Shown in Fig.~\ref{Fig-e906} are the predictions for the DY cross section
ratios at $E_{lab}=120$ GeV in the Fermilab's E906 experiment \cite{ref:e906,ref:e906web}
with invariant dilepton mass $M=4.5$ GeV. At this lower beam proton energy, the target
parton momentum fraction $x$ is large for moderately large beam parton fractional momentum $x^{\prime}$.
The effect of parton shadowing is indeed small as shown by the dashed line in Fig.~\ref{Fig-e906}. On the other hand,
the energy loss effect induced by multiple scattering is significant and the dominant cause for the DY suppression
shown by the shaded bands in Fig. \ref{Fig-e906}. Therefore, the E906 experiment can provide an unambiguous
measurement of the effect of initial state energy loss in DY cross section. This is consistent with another recent 
predictions of the energy loss effects at the E906 experiment \cite{Ivan}.

\section{Summary}
\label{sum}

In this paper, we have calculated the differential cross section of
the Drell-Yan dilepton production in hadron-nucleus collisions in
the lowest order of pQCD, including the effect of initial multiple
parton scattering and induced gluon radiation within the framework
of generalized collinear factorization. In nuclear medium, multiple
parton scattering leads to energy loss of the beam parton which has
a quadratic nuclear size dependence because of the LPM interference
effect. Within the leading logarithmic approximation, we express the
effect of parton energy loss in terms of an effective modified beam
quark distribution function. The modification depends on the
twist-four matrix elements which can be related to the normal
twist-2 parton distributions and jet transport parameter $\hat{q}$.
Using the values of $\hat{q}$ determined from nuclear DIS data by
the HERMES experiment \cite{Wang:2009qb}, we found that the effect
of initial beam parton energy loss is quite small for large
invariant mass of DY lepton pairs in $p+A$ collision at high-energies, except in the large $x^{\prime}$
region and for heavy nuclear targets. The Fermilab E866 experimental
data are consistent with the effect of the nuclear shadowing of
target parton distributions as given by the EPS08 parametrization.
However, the effect of beam parton energy loss becomes significant
for lower beam proton energy $E_{lab}=120$ GeV at the Fermilab E906 experiment, especially
at large $x^{\prime}$ and with heavy nuclear targets. For
parameterization with improved accuracy of nuclear parton
distribution functions through consistent global fitting of
experiment data, one should also include the effect of initial beam
parton energy loss in the DY process.

\begin{center}
{\bf Acknowledgments}
\end{center}
This work is supported by the NSFC of China under Projects Nos.
10825523, 10875025 and MOST of China under project No. 2008CB317106,
and by MOE and SAFEA of China under project No. PITDU-B08033, and by
the Director, Office of Energy Research, Office of High Energy and
Nuclear Physics, Divisions of Nuclear Physics, of the U.S.
Department of Energy under Contract No. DE-AC02-05CH11231. The
research of Y.G. was supported by the Natural Sciences and
Engineering Research Council of Canada.

\appendix
\begin{center}
{\bf Appendix}
\end{center}

In this appendix we list the complete results for annihilation-like
and Compton-like processes, we also provide an cross-check of the
double-hard part through partonic quark-gluon and gluon-gluon
scattering.

\section{complete results in annihilation-like process}
\label{appa}
There are total 9 diagrams for central-cut, and 7 diagrams for
left- and right-cut each. The central-cut diagrams are shown in Fig.
\ref{Fig-DA} and their contributions are:
\begin{eqnarray}
\overline{H}_{A11}&=&\alpha_s^2\alpha_{em}e_q^2C_F\frac{1+z^2}{1-z}\frac{1}{(\vec{q}_T-z\vec{k}_T)^2}\frac{8(2\pi)^2}{N_c^2}\overline{I}_{A11},\\
\nonumber
\overline{I}_{A11}&=&e^{i(x_B+x_t+x_C-x_D)p^+y^-}e^{ix_Dp^+(y_1^--y_2^-)}\theta(-y_2^-)\theta(y^--y_1^-)\\
&\times&\left[1-e^{-i(x_t+x_C-x_D)p^+(y^--y^-_1)}\right]\left[(1-e^{-i(x_t+x_C-x_D)p^+y^-_2}\right]\\
\overline{H}_{A22}&=&\alpha_s^2\alpha_{em}e_q^2C_F\frac{1+z^2}{1-z}\frac{1}{(\vec{q}_T-\vec{k}_T)^2}\frac{8(2\pi)^2}{N_c^2}\overline{I}_{A22},\\
\overline{I}_{A22}&=&e^{ix_Bp^+y^-}e^{i(x_t+x_C)p^+y_1^-}e^{-i(x_t+x_C)p^+y_2^-}\theta(-y_2^-)\theta(y^--y_1^-)\\
\overline{H}_{A33}&=&\alpha_s^2\alpha_{em}e_q^2C_A\frac{1+z^2}{1-z}\frac{1}{{q}_T^2}
\frac{8(2\pi)^2}{N_c^2}\overline{I}_{A33},\\
\nonumber
\overline{I}_{A33}&=&e^{ix_Bp^+y^-}e^{i(x_t+x_C)p^+y_1^-}e^{-i(x_t+x_C)p^+y_2^-}\theta(-y_2^-)\theta(y^--y_1^-);\\
\overline{H}_{A12}&=&\alpha_s^2\alpha_{em}e_q^2\left(C_F-\frac{C_A}{2}\right)\frac{1+z^2}{1-z}
\frac{[q_T^2-(1+z)\vec{k}_T\cdot
\vec{q}_T+zk_T^2]}{(\vec{q}_T-\vec{k}_T)^2(\vec{q}_T-z\vec{k}_T)^2}\frac{8(2\pi)^2}{N_c^2}\overline{I}_{A12},\\
\nonumber
\overline{I}_{A12}&=&e^{i(x_B+x_t+x_C-x_D)p^+y^-}e^{ix_Dp^+(y_1^--y_2^-)}\theta(-y_2^-)\theta(y^--y_1^-)\\
&\times&\left[1-e^{-i(x_t+x_C-x_D)p^+(y^--y^-_1)}\right]e^{-i(x_t+x_C-x_D)p^+y^-_2}\\
\overline{H}_{A13}&=&\alpha_s^2\alpha_{em}e_q^2\frac{C_A}{2}\frac{1+z^2}{1-z}
\frac{q_T^2-z\vec{k}_T\cdot
\vec{q}_T}{q_T^2(\vec{q}_T-z\vec{k}_T)^2}\frac{8(2\pi)^2}{N_c^2}\overline{I}_{A13},\\
\nonumber
\overline{I}_{A13}&=&e^{i(x_B+x_t+x_C-x_D)p^+y^-}e^{ix_Dp^+(y_1^--y_2^-)}\theta(-y_2^-)\theta(y^--y_1^-)\\
&\times&\left[1-e^{-i(x_t+x_C-x_D)p^+(y^--y^-_1)}\right]e^{-i(x_t+x_C-x_D)p^+y^-_2}\\
\overline{H}_{A21}&=&\alpha_s^2\alpha_{em}e_q^2\left(C_F-\frac{C_A}{2}\right)\frac{1+z^2}{1-z}
\frac{q_T^2-(1+z)\vec{k}_T\cdot
\vec{q}_T+zk_T^2}{(\vec{q}_T-\vec{k}_T)^2(\vec{q}_T-z\vec{k}_T)^2}\frac{8(2\pi)^2}{N_c^2}\overline{I}_{A21},\\
\nonumber
\overline{I}_{A21}&=&e^{i(x_B+x_t+x_C-x_D)p^+y^-}e^{ix_Dp^+(y_1^--y_2^-)}\theta(-y_2^-)\theta(y^--y_1^-)\\
&\times&e^{-i(x_t+x_C-x_D)p^+(y^--y^-_1)}\left[(1-e^{-i(x_t+x_C-x_D)p^+y^-_2}\right]\\
\overline{H}_{A23}&=&\alpha_s^2\alpha_{em}e_q^2\frac{C_A}{2}\frac{1+z^2}{1-z}
\frac{q_T^2-\vec{k}_T\cdot \vec{q}_T}{q_T^2(\vec{q}_T-\vec{k}_T)^2}\frac{8(2\pi)^2}{N_c^2}\overline{I}_{A23},\\
\overline{I}_{A23}&=&-e^{ix_Bp^+y^-}e^{i(x_t+x_C)p^+y_1^-}e^{-i(x_t+x_C)p^+y_2^-}\theta(-y_2^-)\theta(y^--y_1^-)\\
\overline{H}_{A31}&=&\alpha_s^2\alpha_{em}e_q^2\frac{C_A}{2}\frac{1+z^2}{1-z}
\frac{q_T^2-z\vec{k}_T\cdot \vec{q}_T}{q_T^2(\vec{q}_T-z\vec{k}_T)^2}\frac{8(2\pi)^2}{N_c^2}\overline{I}_{A31},\\
\nonumber
\overline{I}_{A31}&=&e^{i(x_B+x_t+x_C-x_D)p^+y^-}e^{ix_Dp^+(y_1^--y_2^-)}\theta(-y_2^-)\theta(y^--y_1^-)\\
&\times&e^{-i(x_t+x_C-x_D)p^+(y^--y^-_1)}\left[(1-e^{-i(x_t+x_C-x_D)p^+y^-_2}\right]\\
\overline{H}_{A32}&=&\alpha_s^2\alpha_{em}e_q^2\frac{C_A}{2}\frac{1+z^2}{1-z}
\frac{q_T^2-\vec{k}_T\cdot \vec{q}_T}{q_T^2(\vec{q}_T-\vec{k}_T)^2}\frac{8(2\pi)^2}{N_c^2}\overline{I}_{A32},\\
\overline{I}_{A32}&=&-e^{ix_Bp^+y^-}e^{i(x_t+x_C)p^+y_1^-}e^{-i(x_t+x_C)p^+y_2^-}\theta(-y_2^-)\theta(y^--y_1^-).
\end{eqnarray}

\begin{figure}[h]
\begin{center}
\includegraphics[width=15.0cm]{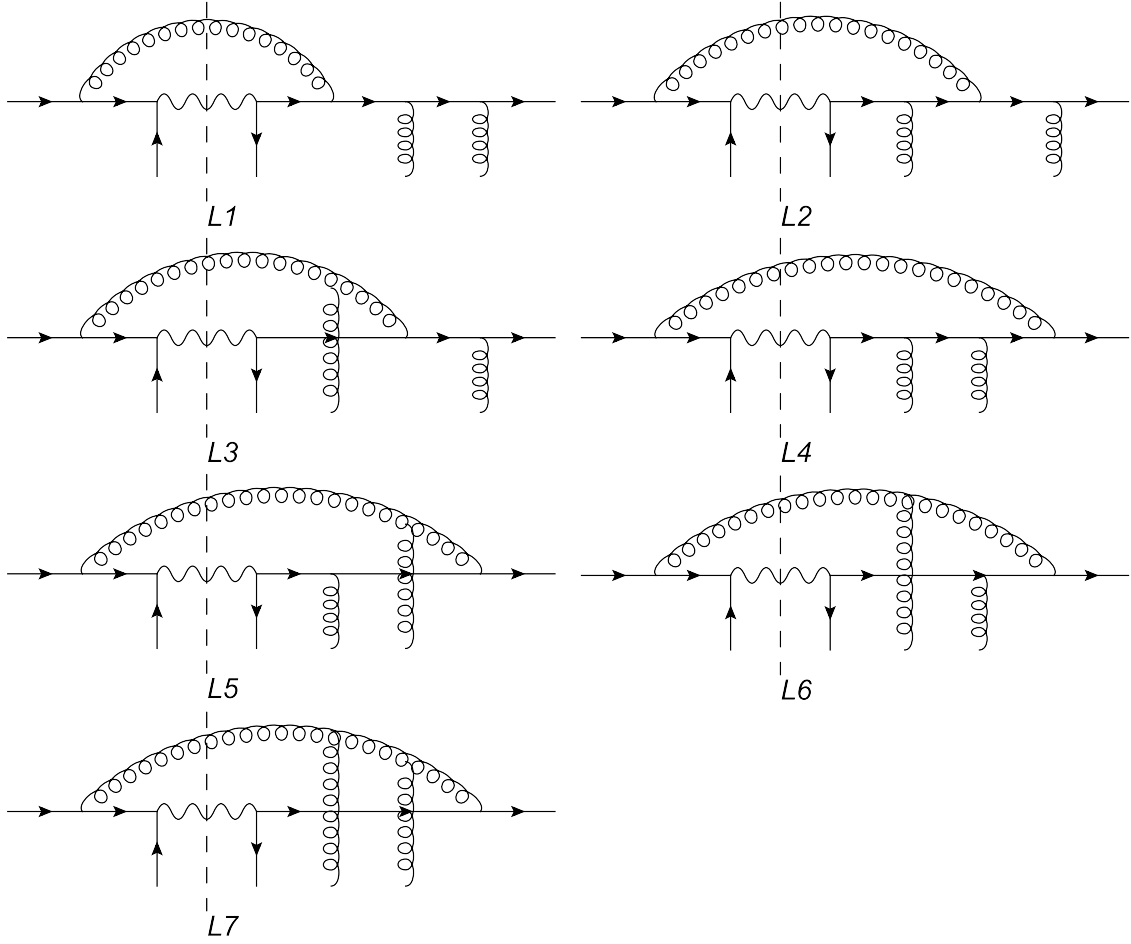}
\centerline{\parbox{15cm}{\caption{\label{Fig-left1}\small Left-cut
diagrams in annihilation-like process.}}}
\end{center}
\end{figure}
All the left-cut diagrams in annihilation-like process are shown in
Fig. \ref{Fig-left1} and their contributions are:
\begin{eqnarray}
\overline{H}_{AL1}&=&\alpha_s^2\alpha_{em}e_q^2C_F\frac{1+z^2}{1-z}\frac{1}{q_T^2}\frac{8(2\pi)^2}{N_c^2}\overline{I}_{L1},\\
\overline{I}_{AL1}&=&-e^{i(x_B+x_t)p^+y^-}e^{-ix_Dp^+(y_1^--y_2^-)}\left[1-e^{-ix_tp^+y_2^-}\right]\theta(-y_2^-)\theta(y_2^--y_1^-);\\
\overline{H}_{AL2}&=&\alpha_s^2\alpha_{em}e_q^2\left(C_F-\frac{C_A}{2}\right)\frac{1+z^2}{1-z}
\frac{q_T^2-(1-z)\vec{k}_T\cdot
\vec{q}_T}{q_T^2[\vec{q}_T-(1-z)\vec{k}_T]^2}
\frac{8(2\pi)^2}{N_c^2}\overline{I}_{L2},\\
\nonumber
\overline{I}_{AL2}&=&e^{i(x_B+x_t)p^+y^-}e^{-ix_Dp^+(y_1^--y_2^-)}\theta(-y_2^-)\theta(y_2^--y_1^-);\\
&\times& e^{-ix_tp^+y_2^-}\left[e^{i(x_E-x_t+x_D)p^+(y_1^--y_2^-)}-1\right];\\
\overline{H}_{AL3}&=&\alpha_s^2\alpha_{em}e_q^2\frac{C_A}{2}\frac{1+z^2}{1-z}
\frac{q_T^2+z\vec{k}_T\cdot
\vec{q}_T}{q_T^2(\vec{q}_T+z\vec{k}_T)^2}
\frac{8(2\pi)^2}{N_c^2}\overline{I}_{L3},\\
\nonumber
\overline{I}_{AL3}&=&e^{i(x_B+x_t)p^+y^-}e^{-ix_Dp^+(y_1^--y_2^-)}\theta(-y_2^-)\theta(y_2^--y_1^-)\\
&\times&e^{-ix_tp^+y_2^-}\left[e^{i(-x_F-x_t+x_D)p^+(y_1^--y_2^-)}-1\right];\\
\overline{H}_{AL4}&=&\alpha_s^2\alpha_{em}e_q^2C_F\frac{1+z^2}{1-z}\frac{1}{q_T^2}
\frac{8(2\pi)^2}{N_c^2}\overline{I}_{L4},\\
\overline{I}_{AL4}&=&-e^{i(x_B+x_t)p^+y^-}e^{i(x_E-x_t)p^+y_1^-}e^{-ix_Ep^+y_2^-}\theta(-y_2^-)\theta(y_2^--y_1^-);\\
\overline{H}_{AL5}&=&\alpha_s^2\alpha_{em}e_q^2\frac{C_A}{2}\frac{1+z^2}{1-z}
\frac{q_T^2-\vec{k}_T\cdot
\vec{q}_T}{(\vec{q}_T-\vec{k}_T)^2q_T^2}\frac{8(2\pi)^2}{N_c^2}\overline{I}_{L5},\\
\overline{I}_{AL5}&=&e^{i(x_B+x_t)p^+y^-}e^{i(x_E-x_t)p^+y_1^-}e^{-ix_Ep^+y_2^-}\theta(-y_2^-)\theta(y_2^--y_1^-);\\
\overline{H}_{AL6}&=&\alpha_s^2\alpha_{em}e_q^2\frac{C_A}{2}\frac{1+z^2}{1-z}
\frac{q_T^2+\vec{k}_T\cdot \vec{q}_T}{q_T^2(\vec{q}_T+\vec{k}_T)^2}\frac{8(2\pi)^2}{N_c^2}\overline{I}_{L6},\\
\overline{I}_{AL6}&=&e^{i(x_B+x_t)p^+y^-}e^{-i(x_F+x_t)p^+y_1^-}e^{ix_Fp^+y_2^-}\theta(-y_2^-)\theta(y_2^--y_1^-);\\
\overline{H}_{AL7}&=&\alpha_s^2\alpha_{em}e_q^2C_A\frac{1+z^2}{1-z}\frac{1}{{q}_T^2}
\frac{8(2\pi)^2}{N_c^2}\overline{I}_{L7},\\
\overline{I}_{AL7}&=&-e^{i(x_B+x_t)p^+y^-}e^{-i(x_F+x_t)p^+y_1^-}e^{ix_Fp^+y_2^-}\theta(-y_2^-)\theta(y_2^--y_1^-).
\end{eqnarray}
where,
\begin{eqnarray}
x_E=\frac{x_B}{Q^2}(2\vec{q}_T\cdot
\vec{k}_T-k_T^2),~~~x_F=\frac{x_B}{Q^2}\frac{z(k_T^2+2\vec{q}_T\cdot
\vec{k}_T)}{1-z}.
\end{eqnarray}

The right-cut diagrams are the complex conjugate of left-cut
diagrams shown in Fig. \ref{Fig-left1}, their contributions are:
\begin{eqnarray}
\overline{H}_{AR1}&=&\alpha_s^2\alpha_{em}e_q^2C_F\frac{1+z^2}{1-z}\frac{1}{q_T^2}\frac{8(2\pi)^2}{N_c^2}\overline{I}_{R1},\\
\overline{I}_{AR1}&=&-e^{i(x_B+x_t)p^+y^-}e^{-ix_Dp^+(y_1^--y_2^-)}\theta(y^--y_1^-)\theta(y_1^--y_2^-)\\
&\times&\left[1-e^{ix_tp^+(y_1^--y_2^-)}
e^{ix_tp^+(y_2^--y^-)}\right];\\
\overline{H}_{AR2}&=&\alpha_s^2\alpha_{em}e_q^2\left(C_F-\frac{C_A}{2}\right)\frac{1+z^2}{1-z}
\frac{q_T^2-(1-z)\vec{k}_T\cdot
\vec{q}_T}{q_T^2[\vec{q}_T-(1-z)\vec{k}_T]^2}
\frac{8(2\pi)^2}{N_c^2}\overline{I}_{R2},\\
\nonumber
\overline{I}_{AR2}&=&e^{i(x_B+x_t)p^+y^-}e^{ix_Ep^+(y_1^--y_2^-)}\theta(y^--y_1^-)\theta(y_1^--y_2^-);\\
&\times&e^{ix_tp^+(y_2^--y^-)}\left[1-e^{i(x_t-x_D-x_E)p^+(y_1^--y_2^-)}\right];\\
\overline{H}_{AR3}&=&\alpha_s^2\alpha_{em}e_q^2\frac{C_A}{2}\frac{1+z^2}{1-z}
\frac{q_T^2+z\vec{k}_T\cdot
\vec{q}_T}{q_T^2(\vec{q}_T+z\vec{k}_T)^2}
\frac{8(2\pi)^2}{N_c^2}\overline{I}_{R3},\\
\nonumber
\overline{I}_{AR3}&=&e^{i(x_B+x_t)p^+y^-}e^{-ix_Fp^+(y_1^--y_2^-)}\theta(y^--y_1^-)\theta(y_1^--y_2^-)\\
&\times&e^{ix_tp^+(y_2^--y^-)}\left[1-e^{i(x_t-x_D+x_F)p^+(y_1^--y_2^-)}\right];\\
\overline{H}_{AR4}&=&\alpha_s^2\alpha_{em}e_q^2C_F\frac{1+z^2}{1-z}\frac{1}{q_T^2}
\frac{8(2\pi)^2}{N_c^2}\overline{I}_{R4},\\
\overline{I}_{AR4}&=&-e^{i(x_B+x_t)p^+y^-}e^{ix_Ep^+(y_1^--y_2^-)}e^{ix_tp^+(y_2^--y^-)}\theta(y^--y_1^-)\theta(y_1^--y_2^-);\\
\overline{H}_{AR5}&=&\alpha_s^2\alpha_{em}e_q^2\frac{C_A}{2}\frac{1+z^2}{1-z}
\frac{q_T^2-\vec{k}_T\cdot
\vec{q}_T}{(\vec{q}_T-\vec{k}_T)^2q_T^2}\frac{8(2\pi)^2}{N_c^2}\overline{I}_{R5},\\
\overline{I}_{AR5}&=&e^{i(x_B+x_t)p^+y^-}e^{ix_Ep^+(y_1^--y_2^-)}e^{ix_tp^+(y_2^--y^-)}\theta(y^--y_1^-)\theta(y_1^--y_2^-);\\
\overline{H}_{AR6}&=&\alpha_s^2\alpha_{em}e_q^2\frac{C_A}{2}\frac{1+z^2}{1-z}
\frac{q_T^2+\vec{k}_T\cdot \vec{q}_T}{q_T^2(\vec{q}_T+\vec{k}_T)^2}\frac{8(2\pi)^2}{N_c^2}\overline{I}_{R6},\\
\overline{I}_{AR6}&=&e^{i(x_B+x_t)p^+y^-}e^{-ix_Fp^+(y_1^--y_2^-)}e^{ix_tp^+(y_2^--y^-)}\theta(y^--y_1^-)\theta(y_1^--y_2^-);\\
\overline{H}_{AR7}&=&\alpha_s^2\alpha_{em}e_q^2C_A\frac{1+z^2}{1-z}\frac{1}{q_T^2}
\frac{8(2\pi)^2}{N_c^2}\overline{I}_{R7},\\
\overline{I}_{AR7}&=&-e^{i(x_B+x_t)p^+y^-}e^{-ix_Fp^+(y_1^--y_2^-)}e^{ix_tp^+(y_2^--y^-)}\theta(y^--y_1^-)\theta(y_1^--y_2^-);
\end{eqnarray}

\section{complete results in Compton-like process}
\label{appb}
There are same amount of diagrams in Compton-like and
annihilation-like processes, a total of 9 diagrams for central-cut,
and 7 diagrams for left- and right-cut each. The central-cut
diagrams are shown in Fig. \ref{Fig-DC} and their contributions are:
\begin{eqnarray}
\overline{H}_{C11}&=&\alpha_s^2\alpha_{em}e_q^2\frac{C_A}{N_c^2-1}\frac{2z^2-2z+1}{(\vec{q}_T-z\vec{k}_T)^2}\frac{8(2\pi)^2}{N_c}\overline{I}_{C11},\\
\overline{H}_{C22}&=&\alpha_s^2\alpha_{em}e_q^2\frac{C_F}{N_c^2-1}\frac{2z^2-2z+1}{(\vec{q}_T-\vec{k}_T)^2}\frac{8(2\pi)^2}{N_c}\overline{I}_{C22},\\
\overline{H}_{C33}&=&\alpha_s^2\alpha_{em}e_q^2\frac{C_F}{N_c^2-1}\frac{2z^2-2z+1}{q_T^2}\frac{8(2\pi)^2}{N_c}\overline{I}_{C33},\\
\overline{H}_{C12}&=&\alpha_s^2\alpha_{em}e_q^2\frac{C_A}{2}\frac{1}{N_c^2-1}(2z^2-2z+1)
\frac{[q_T^2-(1+z)\vec{k}_T\cdot
\vec{q}_T+zk_T^2]}{(\vec{q}_T-\vec{k}_T)^2(\vec{q}_T-z\vec{k}_T)^2}\frac{8(2\pi)^2}{N_c}\overline{I}_{C12},\\
\overline{H}_{C13}&=&\alpha_s^2\alpha_{em}e_q^2\frac{C_A}{2}\frac{1}{N_c^2-1}(2z^2-2z+1)
\frac{q_T^2-z\vec{k}_T\cdot
\vec{q}_T}{q_T^2(\vec{q}_T-z\vec{k}_T)^2}\frac{8(2\pi)^2}{N_c}\overline{I}_{C13},\\
\overline{H}_{C21}&=&\alpha_s^2\alpha_{em}e_q^2\frac{C_A}{2}\frac{1}{N_c^2-1}(2z^2-2z+1)
\frac{q_T^2-(1+z)\vec{k}_T\cdot
\vec{q}_T+zk_T^2}{(\vec{q}_T-\vec{k}_T)^2(\vec{q}_T-z\vec{k}_T)^2}\frac{8(2\pi)^2}{N_c}\overline{I}_{C21},\\
\overline{H}_{C23}&=&\alpha_s^2\alpha_{em}e_q^2\left(C_F-\frac{C_A}{2}\right)\frac{1}{N_c^2-1}(2z^2-2z+1)
\frac{q_T^2-\vec{k}_T\cdot \vec{q}_T}{q_T^2(\vec{q}_T-\vec{k}_T)^2}\frac{8(2\pi)^2}{N_c}\overline{I}_{C23},\\
\overline{H}_{C31}&=&\alpha_s^2\alpha_{em}e_q^2\frac{C_A}{2}\frac{1}{N_c^2-1}(2z^2-2z+1)
\frac{q_T^2-z\vec{k}_T\cdot \vec{q}_T}{q_T^2(\vec{q}_T-z\vec{k}_T)^2}\frac{8(2\pi)^2}{N_c}\overline{I}_{C31},\\
\overline{H}_{C32}&=&\alpha_s^2\alpha_{em}e_q^2\left(C_F-\frac{C_A}{2}\right)\frac{1}{N_c^2-1}(2z^2-2z+1)
\frac{q_T^2-\vec{k}_T\cdot
\vec{q}_T}{q_T^2(\vec{q}_T-\vec{k}_T)^2}\frac{8(2\pi)^2}{N_c}\overline{I}_{C32},
\end{eqnarray}
and
\begin{eqnarray}
\overline{I}_{Cij}=\overline{I}_{Aij}.
\end{eqnarray}

\begin{figure}[h]
\begin{center}
\includegraphics[width=15.0cm]{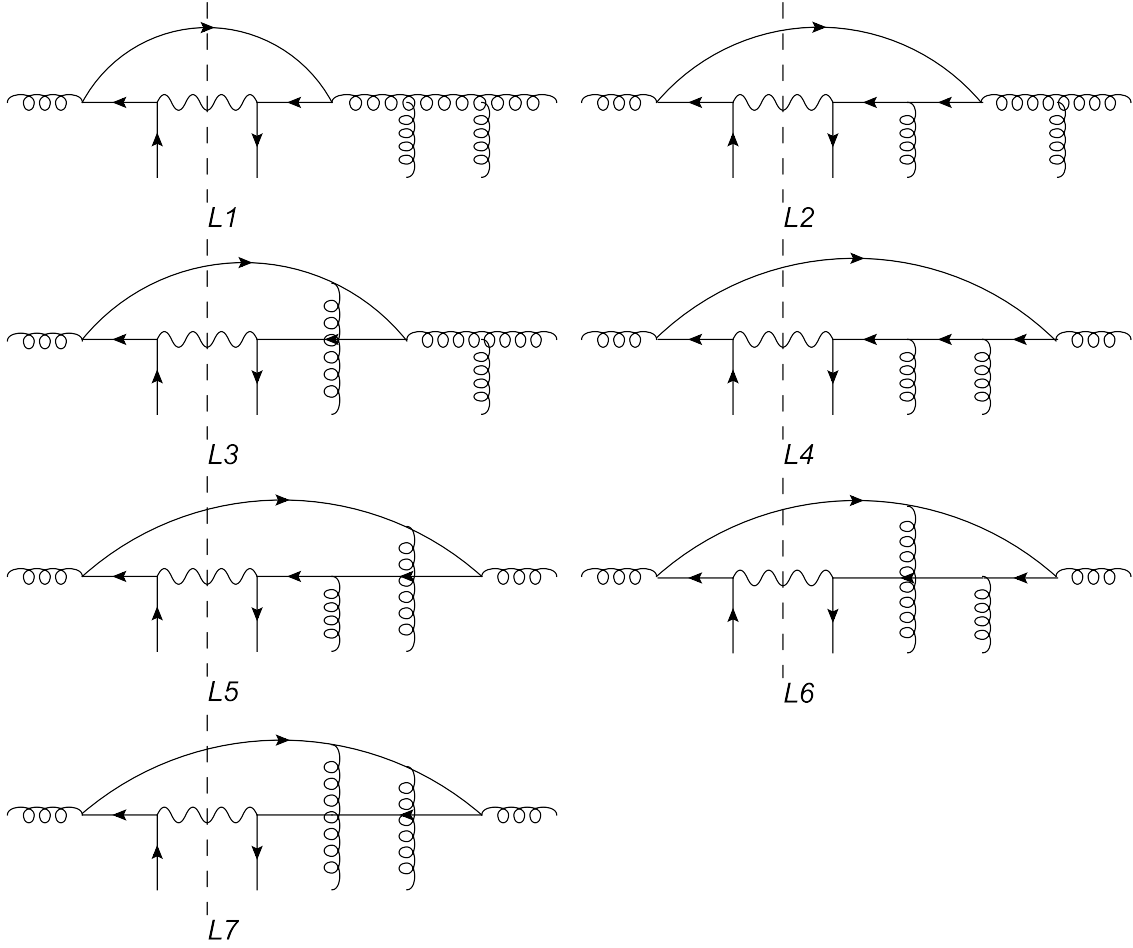}
\centerline{\parbox{15cm}{\caption{\label{Fig-left3}\small Left-cut
diagrams in compton-like process.}}}
\end{center}
\end{figure}
The left-cut diagrams in Compton-like processes are shown in Fig.
\ref{Fig-left3}. Their contributions are:
\begin{eqnarray}
\overline{H}_{CL1}&=&\alpha_s^2\alpha_{em}e_q^2\frac{C_A}{N_c^2-1}(2z^2-2z+1)\frac{1}{q_T^2}\frac{8(2\pi)^2}{N_c}\overline{I}_{L1},\\
\overline{I}_{CL1}&=&-e^{i(x_B+x_t)p^+y^-}e^{-ix_Dp^+(y_1^--y_2^-)}\left[1-e^{-ix_tp^+y_2^-}\right]\theta(-y_2^-)\theta(y_2^--y_1^-);\\
\overline{H}_{CL2}&=&\alpha_s^2\alpha_{em}e_q^2\frac{C_A}{2(N_c^2-1)}(2z^2-2z+1)
\frac{q_T^2-(1-z)\vec{k}_T\cdot
\vec{q}_T}{q_T^2[\vec{q}_T-(1-z)\vec{k}_T]^2}
\frac{8(2\pi)^2}{N_c}\overline{I}_{L2},\\
\nonumber
\overline{I}_{CL2}&=&e^{i(x_B+x_t)p^+y^-}e^{-ix_Dp^+(y_1^--y_2^-)}\theta(-y_2^-)\theta(y_2^--y_1^-);\\
&\times& e^{-ix_tp^+y_2^-}\left[e^{i(x_E-x_t+x_D)p^+(y_1^--y_2^-)}-1\right];\\
\overline{H}_{CL3}&=&\alpha_s^2\alpha_{em}e_q^2\frac{C_A}{2(N_c^2-1)}(2z^2-2z+1)
\frac{q_T^2+z\vec{k}_T\cdot
\vec{q}_T}{q_T^2(\vec{q}_T+z\vec{k}_T)^2}
\frac{8(2\pi)^2}{N_c}\overline{I}_{L3},\\
\nonumber
\overline{I}_{CL3}&=&e^{i(x_B+x_t)p^+y^-}e^{-ix_Dp^+(y_1^--y_2^-)}\theta(-y_2^-)\theta(y_2^--y_1^-)\\
&\times&e^{-ix_tp^+y_2^-}\left[e^{i(-x_F-x_t+x_D)p^+(y_1^--y_2^-)}-1\right];\\
\overline{H}_{CL4}&=&\alpha_s^2\alpha_{em}e_q^2\frac{C_F}{N_c^2-1}(2z^2-2z+1)\frac{1}{q_T^2}
\frac{8(2\pi)^2}{N_c}\overline{I}_{L4},\\
\overline{I}_{CL4}&=&-e^{i(x_B+x_t)p^+y^-}e^{i(x_E-x_t)p^+y_1^-}e^{-ix_Ep^+y_2^-}\theta(-y_2^-)\theta(y_2^--y_1^-);\\
\overline{H}_{CL5}&=&\alpha_s^2\alpha_{em}e_q^2\frac{C_F-C_A/2}{N_c^2-1}(2z^2-2z+1)
\frac{q_T^2-\vec{k}_T\cdot
\vec{q}_T}{(\vec{q}_T-\vec{k}_T)^2q_T^2}\frac{8(2\pi)^2}{N_c}\overline{I}_{L5},\\
\overline{I}_{CL5}&=&e^{i(x_B+x_t)p^+y^-}e^{i(x_E-x_t)p^+y_1^-}e^{-ix_Ep^+y_2^-}\theta(-y_2^-)\theta(y_2^--y_1^-);\\
\overline{H}_{CL6}&=&\alpha_s^2\alpha_{em}e_q^2\frac{C_F-C_A/2}{N_c^2-1}(2z^2-2z+1)
\frac{q_T^2+\vec{k}_T\cdot \vec{q}_T}{q_T^2(\vec{q}_T+\vec{k}_T)^2}\frac{8(2\pi)^2}{N_c}\overline{I}_{L6},\\
\overline{I}_{CL6}&=&e^{i(x_B+x_t)p^+y^-}e^{-i(x_F+x_t)p^+y_1^-}e^{ix_Fp^+y_2^-}\theta(-y_2^-)\theta(y_2^--y_1^-);\\
\overline{H}_{CL7}&=&\alpha_s^2\alpha_{em}e_q^2\frac{C_F}{N_c^2-1}(2z^2-2z+1)\frac{1}{q_T^2}
\frac{8(2\pi)^2}{N_c}\overline{I}_{L7},\\
\overline{I}_{CL7}&=&-e^{i(x_B+x_t)p^+y^-}e^{-i(x_F+x_t)p^+y_1^-}e^{ix_Fp^+y_2^-}\theta(-y_2^-)\theta(y_2^--y_1^-).
\end{eqnarray}

The right-cut diagrams are the complex conjugate of left-cut
diagrams shown in Fig. \ref{Fig-left3} and the results are:
\begin{eqnarray}
\overline{H}_{CR1}&=&\alpha_s^2\alpha_{em}e_q^2\frac{C_A}{N_c^2-1}(2z^2-2z+1)\frac{1}{q_T^2}\frac{8(2\pi)^2}{N_c}\overline{I}_{R1},\\
\overline{I}_{CR1}&=&-e^{i(x_B+x_t)p^+y^-}e^{-ix_Dp^+(y_1^--y_2^-)}\theta(y^--y_1^-)\theta(y_1^--y_2^-)\\
&\times&\left[1-e^{ix_tp^+(y_1^--y_2^-)}
e^{ix_tp^+(y_2^--y^-)}\right];\\
\overline{H}_{CR2}&=&\alpha_s^2\alpha_{em}e_q^2\frac{C_A}{2(N_c^2-1)}(2z^2-2z+1)
\frac{q_T^2-(1-z)\vec{k}_T\cdot
\vec{q}_T}{q_T^2[\vec{q}_T-(1-z)\vec{k}_T]^2}
\frac{8(2\pi)^2}{N_c}\overline{I}_{R2},\\
\nonumber
\overline{I}_{CR2}&=&e^{i(x_B+x_t)p^+y^-}e^{ix_Ep^+(y_1^--y_2^-)}\theta(y^--y_1^-)\theta(y_1^--y_2^-);\\
&\times&e^{ix_tp^+(y_2^--y^-)}\left[1-e^{i(x_t-x_D-x_E)p^+(y_1^--y_2^-)}\right];\\
\overline{H}_{CR3}&=&\alpha_s^2\alpha_{em}e_q^2\frac{C_A}{2(N_c^2-1)}(2z^2-2z+1)
\frac{q_T^2+z\vec{k}_T\cdot
\vec{q}_T}{q_T^2(\vec{q}_T+z\vec{k}_T)^2}
\frac{8(2\pi)^2}{N_c}\overline{I}_{R3},\\
\nonumber
\overline{I}_{CR3}&=&e^{i(x_B+x_t)p^+y^-}e^{-ix_Fp^+(y_1^--y_2^-)}\theta(y^--y_1^-)\theta(y_1^--y_2^-)\\
&\times&e^{ix_tp^+(y_2^--y^-)}\left[1-e^{i(x_t-x_D+x_F)p^+(y_1^--y_2^-)}\right];\\
\overline{H}_{CR4}&=&\alpha_s^2\alpha_{em}e_q^2\frac{C_F}{N_c^2-1}(2z^2-2z+1)\frac{1}{q_T^2}
\frac{8(2\pi)^2}{N_c}\overline{I}_{R4},\\
\overline{I}_{CR4}&=&-e^{i(x_B+x_t)p^+y^-}e^{ix_Ep^+(y_1^--y_2^-)}e^{ix_tp^+(y_2^--y^-)}\theta(y^--y_1^-)\theta(y_1^--y_2^-);\\
\overline{H}_{CR5}&=&\alpha_s^2\alpha_{em}e_q^2\frac{C_F-C_A/2}{N_c^2-1}(2z^2-2z+1)
\frac{q_T^2-\vec{k}_T\cdot
\vec{q}_T}{(\vec{q}_T-\vec{k}_T)^2q_T^2}\frac{8(2\pi)^2}{N_c}\overline{I}_{R5},\\
\overline{I}_{CR5}&=&e^{i(x_B+x_t)p^+y^-}e^{ix_Ep^+(y_1^--y_2^-)}e^{ix_tp^+(y_2^--y^-)}\theta(y^--y_1^-)\theta(y_1^--y_2^-);\\
\overline{H}_{CR6}&=&\alpha_s^2\alpha_{em}e_q^2\frac{C_F-C_A/2}{N_c^2-1}(2z^2-2z+1)
\frac{q_T^2+\vec{k}_T\cdot \vec{q}_T}{q_T^2(\vec{q}_T+\vec{k}_T)^2}\frac{8(2\pi)^2}{N_c}\overline{I}_{R6},\\
\overline{I}_{CR6}&=&e^{i(x_B+x_t)p^+y^-}e^{-ix_Fp^+(y_1^--y_2^-)}e^{ix_tp^+(y_2^--y^-)}\theta(y^--y_1^-)\theta(y_1^--y_2^-);\\
\overline{H}_{CR7}&=&\alpha_s^2\alpha_{em}e_q^2\frac{C_F}{N_c^2-1}(2z^2-2z+1)\frac{1}{q_T^2}
\frac{8(2\pi)^2}{N_c}\overline{I}_{R7},\\
\overline{I}_{CR7}&=&-e^{i(x_B+x_t)p^+y^-}e^{-ix_Fp^+(y_1^--y_2^-)}e^{ix_tp^+(y_2^--y^-)}\theta(y^--y_1^-)\theta(y_1^--y_2^-);
\end{eqnarray}

\section{Cross-check of double hard processes}
\label{appc} One can extract the double-hard subprocesses in
annihilation-like and Compton-like process from Eq. (\ref{Eq DAR})
and Eq. (\ref{Eq DC}),

\begin{eqnarray}
\nonumber \frac{d\sigma_{hA \rightarrow \gamma^{*}}^{AHH}}{dQ^2dx'}
&=&\sum_q\int dxH_0(x,p,q)
\int_0^{\mu^2}\frac{dq_T^2}{q_T^4}\frac{\alpha_s}{2\pi}\int_{x'}^1\frac{d\xi}{\xi}f_{q/h}(\xi)
\frac{1+z^2}{1-z}\frac{2\pi\alpha_s}{N_c}\\
&\times&\left[C_Az+C_F(1-z)^2\right]T_{HH}(x,x_t),\\
\nonumber\frac{d\sigma_{hA \rightarrow \gamma^{*}}^{CHH}}{dQ^2dx'}
&=&\sum_q\int dxH_0(x,p,q)
\int_0^{\mu^2}\frac{dq_T^2}{q_T^4}\frac{\alpha_s}{2\pi}\int_{x'}^1\frac{d\xi}{\xi}f_{g/h}(\xi)
\left[z^2+(1-z)^2\right]\frac{2\pi\alpha_s}{N_c^2-1}\\
&\times&\left[C_F-C_Az(1-z)\right]T_{HH}(x,x_t),
\end{eqnarray}
where,
\begin{eqnarray}
\label{Eq matrix element-HH} \nonumber
T_{HH}(x,x_t)&=&\int\frac{dy^-}{2\pi}dy_1^-dy_2^-e^{ixp^+y^-}e^{ix_tp^+(y_1^--y_2^-)}\\
&\times&\frac{1}{2}\langle
A|F^+_\alpha(y_2^-)\bar{\psi}_q(0)\gamma^+\psi_q(y^-)F^{+\alpha}(y_1^-)|A\rangle
\theta(-y_2^-)\theta(y^--y_1^-)\\
\nonumber &\approx&\pi\int
dy_N^-\rho_A(y_N)f_{q/A}(x)x_tf_{g/N}(x_t)
\end{eqnarray}
with $\rho_A(y_N)$ being the nucleon density distribution.
A factorized form of gluon-quark correlation is assumed in the above equation.
With this
assumption, one can factorized the double-hard subprocess as,
\begin{eqnarray}
\label{Eq AHH} \nonumber \frac{d\sigma_{hA \rightarrow
\gamma^{*}}^{AHH}}{dQ^2} &=&\sum_q\int dxd\xi
H_0(x,p,q)f_{q/h}(\xi)f_{\bar{q}/A}(x)\int dy_N^-\rho_A(y_N)\\
\nonumber &\times&\pi\alpha_s^2x_tf_{g/N}(x_t)
\frac{1+z^2}{1-z}\left[\frac{C_A}{N_c}z+\frac{C_F}{N_c}(1-z)^2\right]dz\frac{dq_T^2}{q_T^4}\\
&\equiv&\frac{d\sigma^{S^{(0)}}_{hA\rightarrow l^+l^-}}{dQ^2}\int dy_N^-\rho_A(y_N)d\sigma_{qN},\\
\nonumber \label{Eq CHH} \frac{d\sigma_{hA \rightarrow
\gamma^{*}}^{CHH}}{dQ^2} &=&\sum_q\int dxd\xi
H_0(x,p,q)f_{g/h}(\xi)f_{q/A}(x)\int dy_N^-\rho_A(y_N)\\
\nonumber &\times&\pi\alpha_s^2x_tf_{g/N}(x_t)
\left[z^2+(1-z)^2\right]
\left[\frac{C_F}{N_c^2-1}-\frac{C_A}{N_c^2-1}z(1-z)\right]dz\frac{dq_T^2}{q_T^4}\\
&\equiv&\frac{d\sigma^{S^{(0)}}_{hA\rightarrow l^+l^-}}{dQ^2}\int
dy_N^-\rho_A(y_N)d\sigma_{gN}.
\end{eqnarray}
Here $d\sigma^{S^{(0)}}_{hA\rightarrow \gamma^{*}}/dQ^2$ is the Born
cross section defined in Eq. (\ref{Eq LO}). $d\sigma_{qN}$ and $d\sigma_{gN}$ represent
the cross section of the parton-nucleon scattering. Since the gluon is
physical and has finite momentum fraction $x_tp^+$ in the
double-hard process, the higher-twist results have simple and
intuitive partonic interpretation as stated in the above equations.
This comes from the assumption that the twist-4 contributions from
 double scattering can be factorized into two  successive and  isolated
single scatterings. In the following, we rederive the twist-4 contributions from
 double-hard scattering based on the
above assumption \cite{benweiqq}.

Consider the parton-nucleon scattering
\begin{eqnarray}
a(p')+N(p)\rightarrow c(q)+d(k)+X
\end{eqnarray}
where,
\begin{eqnarray}
p'=[0,p'^-,0],~~~~~p=[p^+,0,0],~~~~~q=\left[\frac{q_T^2}{2zp'^-},zp'^-,\vec{q}_T\right].
\end{eqnarray}
The parton nucleon cross section can be written as
\begin{eqnarray}
\label{Eq aNcrosssec}
d\sigma_{aN}&=&\int d\sigma_{ag}f_{g/N}(x)dx\\
&=&\int dxf_{g/N}(x)\frac{g^4}{2s}|M|^2_{ag\rightarrow
cd}(s,t,u)\frac{d^3q}{(2\pi)^32q_0}2\pi\delta[(p'+xp-q)^2]\\
&=&\frac{g^4}{(4\pi)^2}x_tf_{g/N}(x_t)|M|^2_{ag\rightarrow
cd}(s,t,u)\frac{\pi}{s^2}\frac{dz}{z(1-z)}dq_T^2,
\end{eqnarray}
where $x_t=q_T^2/[2p\cdot q(1-z)]$ because of the on-shell
condition of the final parton. The
Mandelstam variables in this partonic process can be recast as
\begin{eqnarray}
s=(p'+xp)^2=\frac{q_T^2}{z(1-z)},~~~t=(q-p')^2=-\frac{q_T^2}{z},~~~u=(q-xp)^2=-\frac{q_T^2}{1-z}.
\end{eqnarray}
In the annihilation-like and Compton-like double scattering
processes, the first hard scattering is the
quark-gluon Compton scattering and gluon-gluon fusion, respectively.
The partonic matrix element for these two processes are
\begin{eqnarray}
\nonumber |M|^2_{qg\rightarrow
qg}(s,t,u)&=&\frac{C_A}{N_c}\frac{s^2+u^2}{t^2}-\frac{C_F}{N_c}\frac{s^2+u^2}{us}\\
&=&\frac{C_A}{N_c}\frac{1+z^2}{(1-z)^2}+\frac{C_F}{N_c}\frac{1+z^2}{z}\\
\nonumber |M|^2_{gg\rightarrow
q\bar{q}}(s,t,u)&=&\frac{C_F}{N_c^2-1}\frac{t^2+u^2}{tu}-\frac{C_A}{N_c^2-1}\frac{t^2+u^2}{s^2}\\
&=&\frac{C_F}{N_c^2-1}\frac{z^2+(1-z)^2}{z(1-z)}-\frac{C_A}{N_c^2-1}[z^2+(1-z)^2].
\end{eqnarray}
Substituting the above equations into Eq. (\ref{Eq aNcrosssec}), one can
obtain the quark-nucleon and gluon-nucleon cross section,
\begin{eqnarray}
d\sigma_{qN}&=&\pi\alpha_s^2x_tf_{g/N}(x_t)\left[\frac{C_A}{N_c}z+\frac{C_F}{N_c}(1-z)^2\right]
\frac{1+z^2}{1-z}dz\frac{dq_T^2}{q_T^4},\\
d\sigma_{gN}&=&\pi\alpha_s^2x_tf_{g/N}(x_t)\left[\frac{C_F}{N_c^2-1}-\frac{C_A}{N_c^2-1}z(1-z)\right]
\left[z^2+(1-z)^2\right]dz\frac{dq_T^2}{q_T^4},
\end{eqnarray}
which are equivalent to the results in Eq. (\ref{Eq AHH}) and Eq.
(\ref{Eq CHH}).


\begin{thebibliography}{99}
\bibitem{note} In a very narrow region where
the momentum fraction carried by the beam parton is very close to
zero, there exists a small enhancement effect due to the momentum
sum rule of the parton distribution function. We will ignore this
region in this paper because the lack of experimental data in this region.
\end{thebibliography}
\end{document}